\documentclass[
    amsmath,
    amssymb,
    reprint,
    aps,
    superscriptaddress,
    prd,
    nofootinbib
]{revtex4-2}
\usepackage{natbib}
\usepackage{graphicx}
\usepackage{times}
\usepackage{rotate}
\usepackage{physics}
\usepackage{hyperref}
\usepackage{xcolor}
\usepackage{amsmath}
\usepackage{xspace}
\usepackage{hyperref}
\usepackage{url} 
\usepackage{macros}
\usepackage{soul}
\usepackage{multirow}
\usepackage{bigstrut}  
\usepackage{appendix}
\usepackage{graphicx}
\usepackage{dcolumn}
\usepackage{bm}
\usepackage{aas_macros}
\usepackage{eso-pic}% http://ctan.org/pkg/eso-pic
\newcommand{\mout}{M_{\textrm{out}}}

\newcommand{\SO}{\mathcal{O}}
\newcommand{\M}{\mathcal{M}}

\newcommand{\mvir}{M_{\textrm{vir}}}
\newcommand{\srm}{S_{\lambda}}
\newcommand{\smout}{S_{\mout}}

\newcommand{\redm}{redMaPPer\xspace}

\newcommand{\omegam}{\Omega_{\mathrm{m}}}

  \newcommand{\DESnumbers}{
  \AddToShipoutPictureBG*{%
    \AtPageUpperLeft{%
      \hspace{0.75\paperwidth}%
      \raisebox{-5.5\baselineskip}{%
        \makebox[0pt][l]{\textnormal{DES-2024-0284}}
  }}}%

  \AddToShipoutPictureBG*{%
    \AtPageUpperLeft{%
      \hspace{0.75\paperwidth}%
      \raisebox{-6.5\baselineskip}{%
        \makebox[0pt][l]{\textnormal{FERMILAB-PUB-24-0765-PPD}}
  }}}%
}
 
\begin{document}

% \reportnum{DES-2024-0284}
% \reportnum{FERMILAB-PUB-24-0765-PPD}
  
 \title{\raisebox{-2\baselineskip}{Improving Galaxy Cluster Selection with the Outskirt Stellar Mass of Galaxies}}
 
\author{Matthew Kwiecien}
\email{matt.kwiecien@proton.me}
\affiliation{Department of Physics, University of California, Santa Cruz, CA 95064, USA}
\affiliation{Santa Cruz Institute for Particle Physics, University of California, Santa Cruz, CA 95064, USA}

\author{Tesla Jeltema}
\affiliation{Department of Physics, University of California, Santa Cruz, CA 95064, USA}
\affiliation{Santa Cruz Institute for Particle Physics, University of California, Santa Cruz, CA 95064, USA}

\author{Alexie Leauthaud}
\affiliation{Department of Astrophysics, University of California, Santa Cruz, CA 95064, USA}
\affiliation{Santa Cruz Institute for Particle Physics, University of California, Santa Cruz, CA 95064, USA}

\author{Song Huang}
\affiliation{Department of Astronomy, Tsinghua University, Beijing, China, 100084}

\author{Eli Rykoff}
\affiliation{Kavli Institute for Particle Astrophysics and Cosmology, P. O. Box 2450, Stanford University, Stanford, CA, USA, 94305}
\affiliation{SLAC National Accelerator Laboratory, Menlo Park, CA 94025,
USA}

\author{Sven Heydenreich}
\affiliation{Department of Astrophysics, University of California, Santa Cruz, CA 95064, USA}

\author{Johannes Lange}
\affiliation{Department of Physics, University of Michigan, Ann Arbor, MI 48109, USA}
\affiliation{Leinweber Center for Theoretical Physics, University of Michigan, Ann Arbor, MI 48109, USA}

\author{Spencer Everett}
\affiliation{Jet Propulsion Laboratory, California Institute of Technology, Pasadena, CA 91109, USA}

\author{Conghao Zhou}
\affiliation{Department of Physics, University of California, Santa Cruz, CA 95064, USA}

\author{Paige Kelly}
\affiliation{Department of Physics and Astronomy, University of California Davis, Davis, CA, 95616, USA}

\author{Yuanyuan Zhang}
\affiliation{NSF’s National Optical-Infrared Astronomy Research Laboratory, 950 Cherry Ave.,Tucson, AZ 85719,USA}

\author{Tae-Hyeon Shin}
\affiliation{Department of Physics and Astronomy, Stony Brook University, Stony Brook, New York, USA}

\author{Jesse Golden-Marx}
\affiliation{School of Physics and Astronomy, University of Nottingham, Nottingham,
NG7 2RD, UK}

\author{J.~L.~Marshall}
\affiliation{George P. and Cynthia Woods Mitchell Institute for Fundamental Physics and Astronomy, and Department of Physics and Astronomy, Texas A\&M University, College Station, TX 77843,  USA}

\author{M.~Aguena}
\affiliation{Laborat\'orio Interinstitucional de e-Astronomia - LIneA, Rua Gal. Jos\'e Cristino 77, Rio de Janeiro, RJ - 20921-400, Brazil}
\author{S.~S.~Allam}
\affiliation{Fermi National Accelerator Laboratory, P. O. Box 500, Batavia, IL 60510, USA}
\author{S.~Bocquet}
\affiliation{University Observatory, Faculty of Physics, Ludwig-Maximilians-Universit\"at, Scheinerstr. 1, 81679 Munich, Germany}
\author{D.~Brooks}
\affiliation{Department of Physics \& Astronomy, University College London, Gower Street, London, WC1E 6BT, UK}
\author{A.~Carnero~Rosell}
\affiliation{Instituto de Astrofisica de Canarias, E-38205 La Laguna, Tenerife, Spain}
\affiliation{Laborat\'orio Interinstitucional de e-Astronomia - LIneA, Rua Gal. Jos\'e Cristino 77, Rio de Janeiro, RJ - 20921-400, Brazil}
\affiliation{Universidad de La Laguna, Dpto. Astrofísica, E-38206 La Laguna, Tenerife, Spain}
\author{J.~Carretero}
\affiliation{Institut de F\'{\i}sica d'Altes Energies (IFAE), The Barcelona Institute of Science and Technology, Campus UAB, 08193 Bellaterra (Barcelona) Spain}
\author{L.~N.~da Costa}
\affiliation{Laborat\'orio Interinstitucional de e-Astronomia - LIneA, Rua Gal. Jos\'e Cristino 77, Rio de Janeiro, RJ - 20921-400, Brazil}
\author{M.~E.~S.~Pereira}
\affiliation{Hamburger Sternwarte, Universit\"{a}t Hamburg, Gojenbergsweg 112, 21029 Hamburg, Germany}
\author{T.~M.~Davis}
\affiliation{School of Mathematics and Physics, University of Queensland,  Brisbane, QLD 4072, Australia}
\author{J.~De~Vicente}
\affiliation{Centro de Investigaciones Energ\'eticas, Medioambientales y Tecnol\'ogicas (CIEMAT), Madrid, Spain}
\author{P.~Doel}
\affiliation{Department of Physics \& Astronomy, University College London, Gower Street, London, WC1E 6BT, UK}
\author{I.~Ferrero}
\affiliation{Institute of Theoretical Astrophysics, University of Oslo. P.O. Box 1029 Blindern, NO-0315 Oslo, Norway}
\author{B.~Flaugher}
\affiliation{Fermi National Accelerator Laboratory, P. O. Box 500, Batavia, IL 60510, USA}
\author{J.~Frieman}
\affiliation{Fermi National Accelerator Laboratory, P. O. Box 500, Batavia, IL 60510, USA}
\affiliation{Kavli Institute for Cosmological Physics, University of Chicago, Chicago, IL 60637, USA}
\author{J.~Garc\'ia-Bellido}
\affiliation{Instituto de Fisica Teorica UAM/CSIC, Universidad Autonoma de Madrid, 28049 Madrid, Spain}
\author{M.~Gatti}
\affiliation{Department of Physics and Astronomy, University of Pennsylvania, Philadelphia, PA 19104, USA}
\author{E.~Gaztanaga}
\affiliation{Institut d'Estudis Espacials de Catalunya (IEEC), 08034 Barcelona, Spain}
\affiliation{Institute of Cosmology and Gravitation, University of Portsmouth, Portsmouth, PO1 3FX, UK}
\affiliation{Institute of Space Sciences (ICE, CSIC),  Campus UAB, Carrer de Can Magrans, s/n,  08193 Barcelona, Spain}
\author{G.~Giannini}
\affiliation{Institut de F\'{\i}sica d'Altes Energies (IFAE), The Barcelona Institute of Science and Technology, Campus UAB, 08193 Bellaterra (Barcelona) Spain}
\affiliation{Kavli Institute for Cosmological Physics, University of Chicago, Chicago, IL 60637, USA}
\author{D.~Gruen}
\affiliation{University Observatory, Faculty of Physics, Ludwig-Maximilians-Universit\"at, Scheinerstr. 1, 81679 Munich, Germany}
\author{R.~A.~Gruendl}
\affiliation{Center for Astrophysical Surveys, National Center for Supercomputing Applications, 1205 West Clark St., Urbana, IL 61801, USA}
\affiliation{Department of Astronomy, University of Illinois at Urbana-Champaign, 1002 W. Green Street, Urbana, IL 61801, USA}
\author{G.~Gutierrez}
\affiliation{Fermi National Accelerator Laboratory, P. O. Box 500, Batavia, IL 60510, USA}
\author{S.~R.~Hinton}
\affiliation{School of Mathematics and Physics, University of Queensland,  Brisbane, QLD 4072, Australia}
\author{D.~L.~Hollowood}
\affiliation{Santa Cruz Institute for Particle Physics, Santa Cruz, CA 95064, USA}
\author{K.~Honscheid}
\affiliation{Center for Cosmology and Astro-Particle Physics, The Ohio State University, Columbus, OH 43210, USA}
\affiliation{Department of Physics, The Ohio State University, Columbus, OH 43210, USA}
\author{D.~J.~James}
\affiliation{Center for Astrophysics $\vert$ Harvard \& Smithsonian, 60 Garden Street, Cambridge, MA 02138, USA}
\author{S.~Lee}
\affiliation{Jet Propulsion Laboratory, California Institute of Technology, 4800 Oak Grove Dr., Pasadena, CA 91109, USA}
\author{R.~Miquel}
\affiliation{Instituci\'o Catalana de Recerca i Estudis Avan\c{c}ats, E-08010 Barcelona, Spain}
\affiliation{Institut de F\'{\i}sica d'Altes Energies (IFAE), The Barcelona Institute of Science and Technology, Campus UAB, 08193 Bellaterra (Barcelona) Spain}
\author{A.~Pieres}
\affiliation{Laborat\'orio Interinstitucional de e-Astronomia - LIneA, Rua Gal. Jos\'e Cristino 77, Rio de Janeiro, RJ - 20921-400, Brazil}
\affiliation{Observat\'orio Nacional, Rua Gal. Jos\'e Cristino 77, Rio de Janeiro, RJ - 20921-400, Brazil}
\author{A.~A.~Plazas~Malag\'on}
\affiliation{Kavli Institute for Particle Astrophysics \& Cosmology, P. O. Box 2450, Stanford University, Stanford, CA 94305, USA}
\affiliation{SLAC National Accelerator Laboratory, Menlo Park, CA 94025, USA}
\author{A.~K.~Romer}
\affiliation{Department of Physics and Astronomy, Pevensey Building, University of Sussex, Brighton, BN1 9QH, UK}
\author{S.~Samuroff}
\affiliation{Department of Physics, Northeastern University, Boston, MA 02115, USA}
\affiliation{Institut de F\'{\i}sica d'Altes Energies (IFAE), The Barcelona Institute of Science and Technology, Campus UAB, 08193 Bellaterra (Barcelona) Spain}
\author{E.~Sanchez}
\affiliation{Centro de Investigaciones Energ\'eticas, Medioambientales y Tecnol\'ogicas (CIEMAT), Madrid, Spain}
\author{B.~Santiago}
\affiliation{Instituto de F\'\i sica, UFRGS, Caixa Postal 15051, Porto Alegre, RS - 91501-970, Brazil}
\affiliation{Laborat\'orio Interinstitucional de e-Astronomia - LIneA, Rua Gal. Jos\'e Cristino 77, Rio de Janeiro, RJ - 20921-400, Brazil}
\author{I.~Sevilla-Noarbe}
\affiliation{Centro de Investigaciones Energ\'eticas, Medioambientales y Tecnol\'ogicas (CIEMAT), Madrid, Spain}
\author{M.~Smith}
\affiliation{Physics Department, Lancaster University, Lancaster, LA1 4YB, UK}
\author{E.~Suchyta}
\affiliation{Computer Science and Mathematics Division, Oak Ridge National Laboratory, Oak Ridge, TN 37831}
\author{M.~E.~C.~Swanson}
\affiliation{Center for Astrophysical Surveys, National Center for Supercomputing Applications, 1205 West Clark St., Urbana, IL 61801, USA}
\author{G.~Tarle}
\affiliation{Department of Physics, University of Michigan, Ann Arbor, MI 48109, USA}
\author{D.~L.~Tucker}
\affiliation{Fermi National Accelerator Laboratory, P. O. Box 500, Batavia, IL 60510, USA}
\author{V.~Vikram}
\affiliation{Argone National Laboratory, 9700 S. Cass Avenue, Lemont, IL 60439, USA}
\author{N.~Weaverdyck}
\affiliation{Department of Astronomy, University of California, Berkeley,  501 Campbell Hall, Berkeley, CA 94720, USA}
\affiliation{Lawrence Berkeley National Laboratory, 1 Cyclotron Road, Berkeley, CA 94720, USA}
\author{P.~Wiseman}
\affiliation{School of Physics and Astronomy, University of Southampton,  Southampton, SO17 1BJ, UK}

% These dates will be filled out by the publisher
\date{Accepted XXX. Received YYY; in original form ZZZ}

\begin{abstract} 

The number density and redshift evolution of optically selected galaxy clusters offer an independent measurement of the amplitude of matter fluctuations, $S_8$. However, recent results have shown that clusters chosen by the \redm algorithm show richness-dependent biases that affect the weak lensing signals and number densities of clusters, increasing uncertainty in the cluster mass calibration and reducing their constraining power. In this work, we evaluate an alternative cluster proxy, outskirt stellar mass, $\mout$, defined as the total stellar mass within a $[50,100]$ $\kpc$ envelope centered on a massive galaxy. This proxy exhibits scatter comparable to \redm richness, $\lambda$, but is less likely to be subject to projection effects. We compare the Dark Energy Survey Year 3 \redm cluster catalog with a $\mout$ selected cluster sample from the Hyper-Suprime Camera survey. We use weak lensing measurements to quantify and compare the scatter of $\mout$ and $\lambda$ with halo mass. Our results show $\mout$ has a scatter consistent with $\lambda$, with a similar halo mass dependence, and that both proxies contain unique information about the underlying halo mass. We find $\lambda$-selected samples introduce features into the measured $\Delta \Sigma$ signal that are not well fit by a log-normal scatter only model, absent in $\mout$ selected samples. Our findings suggest that $\mout$ offers an alternative for cluster selection with more easily calibrated selection biases, at least at the generally lower richnesses probed here. Combining both proxies may yield a mass proxy with a lower scatter and more tractable selection biases, enabling the use of lower mass clusters in cosmology. Finally, we find the scatter and slope in the $\lambda-\mout$ scaling relation to be $0.49 \pm 0.02$ and $0.38 \pm 0.09$.

\end{abstract}
\DESnumbers
\maketitle

\section{Introduction}

Within the $\Lambda$ cold dark matter ($\Lambda$CDM) model, the halo mass function (HMF) predicts the abundance of dark matter halos per unit co-moving volume \citep[e.g.][]{PS1974,ST1999,Jenkins01} with percent level accuracy \citep{Tinker08,Bocquet16}.  The observed number, mass distribution, and redshift evolution of dark matter halos can be used to place stringent constraints on cosmological parameters \citep[e.g.][]{Allen2011, Kravtsov12, Burenin12, Mantz15, Cataneo15}.  However, direct observation of dark matter is currently impossible, so we must rely on mass proxies to probe the HMF. Galaxy clusters are excellent candidates as they emerge from the highest density peaks of the primordial matter distribution \citep{Mantz10, Huterer18}, tracing the high end of the mass function. Clusters can constrain the present-day mean energy density of matter, $\omegam$ \citep{Vikhlinin09}, and the root mean square amplitude of linear mass fluctuations of the early universe smoothed over spheres of $8\,\h\Mpc$ at the present epoch, $\sigma_8$ \citep{PS1974}, through the amplitude, redshift evolution, and shape of the mass function. The precise measurement of $\omegam$ from galaxy clusters is crucial in measuring the mean energy density of dark energy, $\Omega_\Lambda = 1 - \omegam$, and in breaking the degeneracy between the dark energy equation of state $\omega$ and $\omegam$ in Cosmic Microwave Background (CMB) and Type Ia supernova (SNe Ia) analyses \citep{Huterer18}.  

In recent years, there have been hints at a disagreement between late and early time measurements of the parameter $S_8 \equiv \left( \frac{\omegam}{0.3}\right)^{0.5} \sigma_8$ \citep{Handley19, Muir21}, a measure of the in-homogeneity of the universe at $8\h\Mpc$ scales, which is roughly the size of a region that clusters collapse from.  When $S_8$ is calculated using CMB data at $z \sim 1100$ and forward modeled to the present day, a consistently higher value of $S_8$ is predicted \citep{Planck2018, ACTDR4} than what is observed in the local universe when using a combination of galaxy clustering, weak lensing, and cosmic shear \citep{Lange23, DESY3-wl, KiDS21, HSCY3, HSCY3-2, Leauthaud17}.  While the high and low redshift measurements of $S_8$ are in statistical agreement, most local measurements of $S_8$ using weak lensing and galaxy clustering are systematically lower. However, alternative low redshift measurements using CMB lensing are higher and consistent with Planck \citep{ACTDR6}.  This emergent tension could suggest new physics, but it remains to be seen if there are unknown systematics in weak lensing or CMB measurements affecting the results.  Galaxy clusters offer an alternative and independent probe of $S_8$ with very different systematic effects, which could confirm or rule out new physics. 

X-ray emission and the Sunyaev-Zeldovich (SZ) signal are excellent methods for selecting clusters \citep[e.g.][]{deHaan16, Mantz14, Vikhlinin09},  but are limited to the very high mass end of the halo mass function. They are, therefore, limited in number counts and statistical uncertainty, whereas finding clusters in the optical bands \citep{DESY1-clusters} can probe to much lower halo mass and give considerable statistical constraining power.  Finding the number density of optical clusters requires a ``cluster-finder'' that can be used on large-scale sky surveys to identify and count clusters at different redshifts.  The current optical cluster-finder used by many cluster analyses is the red-sequence Matched-filter Probabilistic Percolation (\redm) algorithm, described in \citet{Rykoff16, Rykoff2014}. Historically, a cluster's richness was understood as the count of galaxies within a radius \citep{Zwicky1933}, but \redm provides an optimized richness, $\lambda$, which has a low scatter at fixed halo mass, i.e., $\sigma_{\lambda|\M}$ in the $M_\mathrm{halo}-\lambda$ relation \citep{Rykoff2014}. 

Recent work has shown that \redm introduces an additional selection effect that biases the weak lensing signal of \redm selected clusters high in the outskirts of the weak lensing radial profile \citep{Wu22, Huang2022}, which in turn affects the cluster mass calibration and biases cosmology.  \citet{Wu22, Sunayama2020} have shown that these features originate from the presence of interloping structure along the line of sight, colloquially referred to as ``projection effects''.  The systematic error in lensing and mass estimates from projection effects can be corrected for \citep{DESY1-clusters}, but doing so requires simulations with accurate galaxy populations within clusters. Accurately modeling cluster environments within simulations is challenging for the field \citep{To23, DeRose19} and limits our ability to calibrate this systematic.  The complexity in calibrating these selection biases motivates us to study alternative approaches for cluster finding and halo mass tracers than those provided by the current \redm algorithm. 

We investigate the scatter and selection performance of the outskirt stellar mass proxy, $\mout$, which we define as the total stellar mass within a massive galaxy's $[50,100] \,\kpc$ envelope.  Massive galaxies since $z \sim 2$ are known to follow an ``inside-out'' formation model where the inside or \textit{in-situ} stars 
and the outside or \textit{ex-situ} stars grow through different mechanisms \citep{Oser10, vanDokkum10}. The \textit{in-situ} component that is assembled primarily through the main progenitor, and the \textit{ex-situ} component that grows through dry mergers which are more likely to trace the mass of a dark matter halo \citep{Qu17}.  \citet{Bradshaw20} found in simulations that this \textit{ex-situ} mass traces dark matter with lower scatter at fixed halo mass than total stellar mass.  \citet{Huang2022} used a combination of simulations and stacked excess surface density measurements ($\Delta \Sigma$) to quantify the scatter of aperture stellar mass measurements in different radial apertures using Hyper Suprime-Cam Subaru Strategic Program (HSC-SSP) \citep{Aihara18b} data.  Those authors found that the radial aperture of $50-100\kpc$ resulted in the lowest scatter of all different radial ranges, and the value was comparable to the scatter of \redm richness, $\lambda$ \citep{Huang2022}. $\mout$ in this radial range closely relates to the intracluster light (ICL) in that ICL surrounds the central galaxies of clusters with a radial extent anywhere from $30\kpc$ up to $1\Mpc$ and with the ICL transition region being between $30-80\kpc$ \citep{Iodice16, ZhangICL19, Montes21}.  The $\mout$ measurement is therefore sampling some of the ICL of each cluster.

The constraining power from the forthcoming data from stage IV experiments such as Rubin Observatory's Legacy Survey of Space and Time (LSST) and the Dark Energy Spectroscopic Instrument (DESI) can resolve open questions about tensions in cosmology \citep{LSST19, DESI16}.  However, systematic effects originating from cluster selection limit our conclusions from cluster cosmology.  For example, the Dark Energy Survey (DES) results on $S_8$ using the \redm algorithm \citep{DESY1-clusters} are in much larger tension with Planck than $S_8$ constraints derived from the same data using cosmic shear measurements \citep{DESY1-wl}. However, \citet{DESY1-clusters} suggests the discrepancy resides in un-modeled cluster selection or mass estimation systematics, particularly at low richness.  A new cluster finder and mass proxy with more straightforward systematic effects, such as $\mout$, may allow us to capitalize on the new data sets from LSST, DESI, and additional future surveys more easily.

These findings motivate us to compare the performance of $\mout$ to $\lambda$ as a cluster finder and cluster mass proxy.  We take clusters selected by $\mout$ in the HSC S16A data release and prepared by \citep{Huang2020} and compare them to the DES Y3 \redm cluster catalog using weak gravitational lensing measurements.  In \autoref{sec:data}, we discuss the optical data used in this analysis. In \autoref{sec:theory}, we describe the simulations we use for modeling our data vectors and the theoretical background for weak gravitational lensing. Then, in \autoref{sec:methods}, we explain how we categorize cluster detections within each catalog and the methods used to compare the scatter and selection of both $\mout$ and $\lambda$. In \autoref{sec:results}, we present the results of our visual inspection, match categorization, and weak gravitational lensing measurements. We interpret and discuss the consequences of our findings in  \autoref{sec:discussion}.  We summarize and conclude in \autoref{sec:conclusions}.

We assume $H_0 = 70$ $\km/\mathrm{s}/\Mpc$, $\omegam = 0.3$, and $\Omega_\Lambda = 0.7$. All masses are reported in units of $\log_{10}(M_{\mathrm{vir}} / M_{\odot})$, where $M_{\mathrm{vir}}$ is defined as $M_{\textrm{vir}} = 4\pi r^3_{\textrm{vir}} \rho_{\textrm{crit}} \Delta_{c} / 3$ (\citealp[for detail, see][]{Bryan1998}).

\begin{figure*}

    \centering
    \includegraphics[width=.95\textwidth]{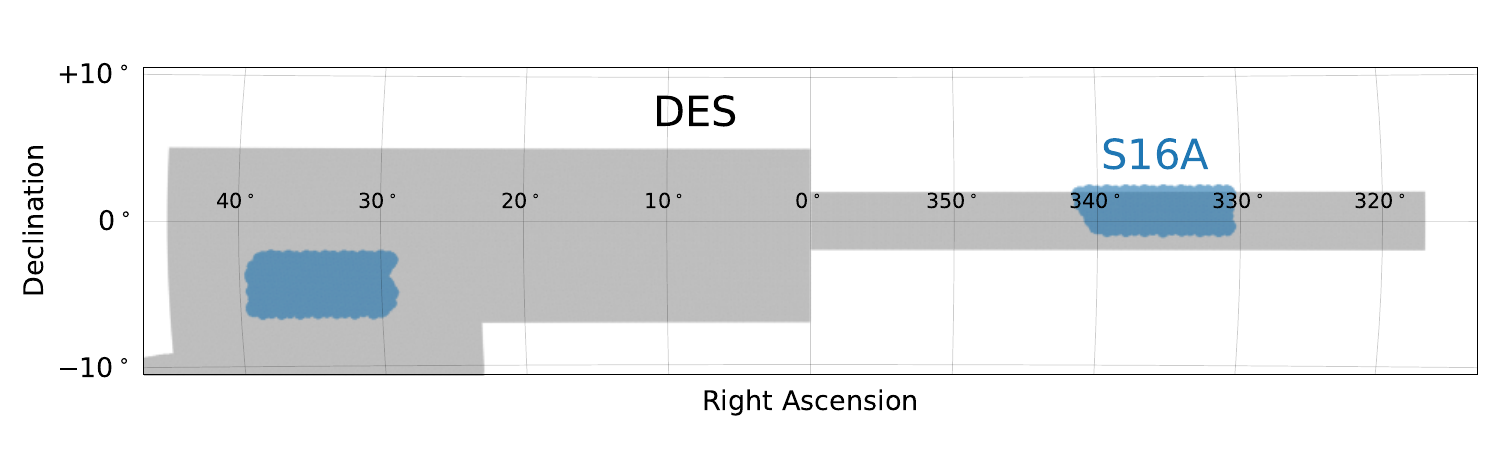}
    \includegraphics[width=.95\textwidth]{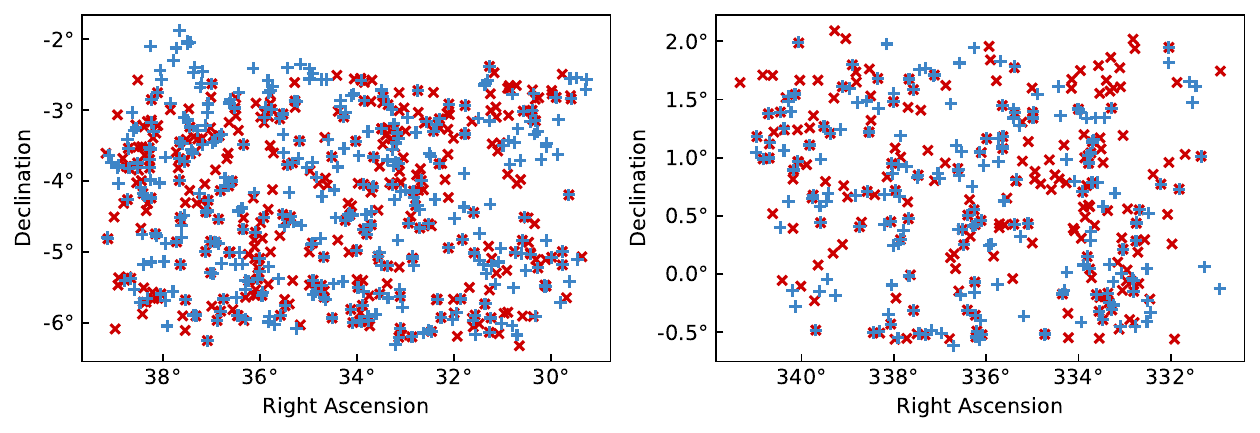}
   
    \caption{Visualization of $\mout$ selected galaxies in HSC data ($\smout$) and \redm clusters in DES Y3 data ($\srm$) after applying all masks and cuts described in \autoref{sec:data}. Top row: the overlap between the two surveys, with the DES footprint in grey and the HSC S16A footprint in blue.  Bottom row: The overlapping regions in detail, with red crosses representing \redm clusters and blue pluses representing $\mout$ selected galaxies.  Left and right plots show the same but for different regions. For the \redm clusters, we only show the \redm central galaxy.  We have 603 \redm clusters and 608 $\mout$ galaxies between $0.19 < z < 0.51$.}
    
    \label{fig:overlap}
    
\end{figure*}

\section{Data}\label{sec:data}

\subsection{HSC survey data}\label{sec:hsc}
The HSC-SSP was a multi-band (\textit{grizY}) imaging survey conducted from 2014-2019 with the Hyper Suprime-Cam on the 8.2m Subaru Telescope \citep{Aihara18}. The HSC-SSP data has a $5\sigma$ point source detection limit of 26.0 mag, which, combined with the 0.168 arcsec pixel scale of the Subaru telescope \citep{Aihara18b}, enables a precise measurement of outer galaxy light, and therefore $\mout$, for individual galaxies. \citep{Huang18}. For our lensing measurements we use $i$-band images from S16A public shape catalog \citep{Mandelbaum18, Mandelbaum18b} which estimates galaxy shapes using the re-Gaussianization algorithm developed by \citet{Hirata03}.  We estimate the photometric redshifts of the HSC galaxies using the \texttt{frankenz} algorithm \citep{Speagle19}. The S16A shape catalog provides estimates for the intrinsic shape dispersion, shape measurement error, and multiplicative shear bias \citep{Mandelbaum18}, and we incorporate these systematic errors plus a correction factor for the photo-$z$ dilution effect following the methodology of \citet{Huang2022} into our lensing analysis.  

We make use of a massive galaxy catalog prepared by \citet{Huang2020}, which uses $\sim 136$ deg$^2$ of data from the \textsc{WIDE} layer of the internal S16A data release of HSC-SSP and consists of galaxies with $\log_{10}(M_{\star}/\msun) > 10.8$. Briefly, the catalog is constructed by initially selecting all galaxies with $i_{\mathrm{CModel}} \leq 22.0$ mag and with redshift $0.19 < z_{\mathrm{best}} < 0.51$, where $z_\mathrm{{best}}$ is a spectroscopic redshift for the galaxy when available, and a photometric redshift otherwise. Then, a custom approach is used to re-measure the total luminosities of each galaxy in this sample.  The custom photometry is done by measuring each galaxy's 1D surface brightness profiles using elliptical isophote fitting, then integrating to obtain the total flux. Five-band spectral energy distribution (SED) fitting is performed using the improved luminosity measurements to obtain estimates of stellar mass.  Lastly, all galaxies with $\log_{10}(M_{\star}/\msun) > 10.8$ are selected, and then their stellar mass within $50-100\kpc$ is measured from this total stellar mass. For additional details, we refer the reader to \citet{Huang18, Huang2020}.

We consider clusters in the redshift range $0.19 < z < 0.51$, which is the limit of the HSC data we use.  We apply the DES Y3A2 Gold footprint mask and Y3A2 Gold foreground bright star masks, except for the 2MASS faint stars mask \citep{Sevilla21}, to the S16A sample to have consistent masking across both samples. Since the creation of the massive galaxy catalog, a new S18A bright star mask that improves on the original S16A bright star mask has been released \citep{Coupon18}, and we apply this updated bright star mask to the HSC data.  Lastly, we use an $\mout \left[\log_{10}\left(M_{\textrm{vir}}/\msun\right)\right] \geq 10.63$ cut to the massive galaxy catalog, roughly corresponding to $\lambda \geq 6$. This cut was selected so that both input catalogs have the same overall number density and reduces the number of galaxies in the $\mout$ catalog from 8,021 to 608.  After applying our cuts and masks, the remaining outer mass galaxies form the sample designated as $\smout$.  We are left with 608 outer-mass selected galaxies in $\smout$.  This sample has a spectroscopic completeness of $88\%$ with 538 galaxies having a spectroscopic redshift ($\zspec$).

\subsection{DES Y3 \redm catalogs}\label{sec:des}

We run \redm v0.8.5\footnote{https://github.com/erykoff/redmapper} (\textsc{python}) \citep{Rykoff2014, Rykoff16} on the DES Y3A2 Gold 2.2.1 data \citep{Sevilla21} to obtain cluster and member catalogs.  We use the \textsc{python} version for consistency when comparing the DES cluster catalog to our \redm \textsc{runcat} runs (described in \autoref{sec:runcat}).  Additionally this is the publicly released version that will be used for both the DES Y6 cluster analysis and LSST. The \redm cluster finder is a matched-filter, red-sequence based iterative algorithm that first calibrates a model of the red sequence as a function of redshift, then uses that model to identify and assign galaxy clusters with a richness, $\lambda$.  The richness mass proxy $\lambda$ is defined as the excess number of red sequence galaxies with luminosity $L > 0.2 L_{\star}$, the characteristic luminosity for a large galaxy \citep{Schechter76}, within the cluster radius defined to be $R_{\lambda} = 1.0 \,\h \Mpc \, (\lambda / 100)^{0.2}$. We use the spectroscopic redshift of the central galaxy when available and the red-sequence redshift for the cluster, $z_\lambda$, otherwise. We then apply the HSC S16A footprint mask and the S18A bright star mask to both \redm cluster and member catalogs. Lastly, we apply a $\lambda \geq 6$ ($M_{\textrm{vir}} \gtrsim 7.2 \times 10^{13}$ M$_{\odot}$) cut on this catalog to obtain our final cluster and member catalogs, reducing the number of galaxy clusters and member galaxies from $820$ and $13770$ to $603$ and $11431$, respectively. We do not make any cuts on \redm centering probabilities, \texttt{P\_CEN}.  After these cuts and masks are applied, the remaining \redm member galaxies form the sample designated $\srm$.  We are left with 603 \redm clusters and 11431 \redm members in the overlap.  The clusters have a spectroscopic completeness of $68 \%$ with 411 having an available $\zspec$, and the members have a spectroscopic completeness of $15.1 \%$ with 1725 having a $\zspec$.  We illustrate the overlap between the clusters and the HSC massive galaxies in \autoref{fig:overlap} and display only the \redm cluster central galaxy for clarity.

\subsection{$M_{out}$ Cluster Catalog}
\label{sec:runcat}
Due to the existence of satellite galaxies, not every galaxy within $\smout$ is the central galaxy of a galaxy cluster.  This may result in multiple galaxies in $\smout$ tracing the same galaxy cluster as a single \redm cluster in $\srm$. We determine which galaxies in $\smout$ are satellite galaxies by using \redm's percolation algorithm.  The \textit{probabilistic percolation} step is outlined in detail in \citep{Rykoff2014}.  We rank order $\smout$ by $\mout$, then run \redm with percolation to compute a $\lambda$ for each galaxy in $\smout$. During this process, \redm will measure $\lambda$ (and $z_\lambda$) at each galaxy.  Galaxies identified as members of a higher ranked $\mout$ centered cluster will be excluded as potential members of other clusters.  If the overlap with another $\mout$ centered cluster masks enough galaxies such that the measured richness will fall below six, the lower ranked $\mout$ cluster is removed from the list.  The lower ranked $\mout$ galaxy will not be assigned a richness $\lambda$ and will contribute to the higher ranked cluster. We use this resultant catalog, the \texttt{runcat} catalog, to determine which galaxies in $\smout$ are satellites.  We designate richnesses measured with this custom \redm mode to be $\lambda_{\mathrm{rc}}$.  A $\mout$ galaxy with no measured richness $\lambda_{\mathrm{rc}}$ is assumed to be a satellite.

Satellite $\mout$ galaxies are removed from all of the results presented in \autoref{sec:results} except for those in \autoref{subsec:results_match} where we discuss the match rate, types of matches, and failure modes in the two catalogs.

In order to make comparisons between $\lambda$ and $\mout$, we need to combine the two data sets.  We take HSC weak gravitational lensing measurements on the samples we construct after combining the data. The methods we use for this and the final samples are described later in \autoref{sec:methods}.  

\section{Simulations \& Theoretical Methods}\label{sec:theory}

\subsection{N-Body Simulations}

We use a combination of MultiDark-Planck (MDPL2) and Small MultiDark Planck (SMDPL) N-Body simulations \citep{Klypin2016} to allow for low mass halos that may scatter into a cluster selection down to $\lambda\sim5$.  The MDPL2 simulation contains $3840^3$ particles within a 1 Gpc$/h$ box and has particle mass resolution of $1.51 \times 10^9 \msun/h$.  We choose snapshot $z_{\textrm{MDPL2}} = 0.364$ and $z_{\textrm{SMDPL}} = 0.404$ to match the mean redshift of $z \sim 0.4$ of the HSC sample.  We populate MDPL2 with mock observables, varying the scatter in observable at fixed mass, $\sigma_{\SO|\M}$, from 0 to $0.65$ dex in $0.01$ dex increments. The mock observables in our simulations are generic quantities, generated by assuming a log-normal scatter with mass, the expectations of which we can then compare to actual observed quantities like richness or $\mout$; we expand on this formalism in \autoref{subsec:sims}. To resolve lower mass halos with large scatter, we combine these simulations with the SMDPL simulations, which also contain $3840^3$ particles except within a 0.4 Gpc$/h$ box.  The particle resolution in the SMDPL simulation goes down to $9.63 \times 10^7 \msun /h$, and we populate this simulation with observables, varying $\sigma_{\SO|\M}$ from $0.65$ to $1.0$ dex in $0.05$ dex increments.  To combine these two simulations, we use the overlap in the scatter at $\sigma_{\SO | \M} = 0.65$ and measure the excess surface density profiles for observables in each simulation and confirm that each profile is consistent with each other.  We refer the reader to \citet{Huang2022} for more details about combining the simulations.

\subsection{Galaxy-galaxy Lensing}
\label{subsec:ggl}

We use galaxy-galaxy lensing, or the mean tangential shear of source (background) galaxies generated by lens (foreground) galaxies, to measure the surface mass density, $\Delta\Sigma$, around each lens.  We briefly summarize the galaxy-galaxy lensing formalism here \citep[see][for review]{Kilbinger15}.

The average (projected) surface density of a galaxy inside a circle of radius $R$ on the sky is given by
\begin{align}
\langle \Sigma(R)\rangle &= \int_{0}^{\infty} \rho\left(\sqrt{R^2+z^2}\right)\, \mathrm{d} z\\
&= \int_{0}^{\infty} \omegam \rho_{\mathrm{c}} \left(1+\xi_{\mathrm{gm}}\left(\sqrt{R^2+z^2}\right)\right)\, \mathrm{d} z,
\end{align}
where $z$ is the distance along the line of sight from the center of the circle, $\omegam$ is the mean matter density of the universe, $\rho_{\mathrm{c}}$ is the critical density of the universe, and $\xi_{\mathrm{gm}}$ is the galaxy-matter cross-correlation function. The excess surface density, $\Delta \Sigma$, is then obtained by taking the area-weighted average surface density within radius $R$ and subtracting the mean surface density at radius $R$. 
\begin{align}
    \Delta \Sigma & \equiv \left<\Sigma(<R)\right> - \overline\Sigma(R),
\end{align}
which removes the background mean matter density contribution.  Thus, the excess surface density, $\Delta \Sigma$, directly measures the excess matter above the background mean matter density around a lens galaxy. 

The induced tangential shear, $\gamma_t$, is the amount in which a foreground mass distribution tangentially distorts the shape of a background galaxy.  For a spherically symmetric lens galaxy, $\gamma_t$ of a source galaxy can be written as
\begin{equation}
    \gamma_t = \frac{\Delta \Sigma}{\Sigma_{c}},
\end{equation}
where $\Sigma_{c}$, the critical surface mass density in physical units is given by
\begin{equation}\label{eq:sigma-c}
    \Sigma_{c} = \frac{c^2}{4\pi G}\frac{r_{S}}{r_{L}r_{LS}} 
\end{equation}
where $r_{S}$, $r_{L}$, $r_{LS}$ are the angular diameter distances to the source galaxy, to the lens galaxy, and the distance between the two \citep{Miralda1991}.  Then, 
\begin{equation}
    \Delta \Sigma = \gamma_t \Sigma_c, 
\end{equation}
demonstrates that with the $\gamma_t$ of source galaxies around a lens galaxy and the associated $\Sigma_c$ value, we can estimate $\Delta \Sigma$ and the mass of a lens galaxy. To measure $\gamma_t$, we use tangential ellipticities of source galaxies because they are a mostly unbiased tracer of $\gamma_t$. 

Our methodology for computing $\Delta \Sigma$ is summarized in \citet{Huang2020} and described in detail in \citet{Lange19, Speagle19, Singh17}.  Briefly, we measure $\Delta \Sigma$ as

\begin{equation}
    \Delta\Sigma_{\textrm{lr}}(R) = \left(\frac{\sum_{i} w_i \gamma_t^i \Sigma_{c}^i}{\sum_{j} w_j}\right)_{l} - \left(\frac{\sum_{i} w_i \gamma_t^i \Sigma_{c}^i}{\sum_{j} w_j}\right)_{r} 
\end{equation}
where the sum over $i, j$  with outer subscript $l$ is over all lens-source pairs and with outer subscript $r$ is over all random-source pairs, $\gamma_t$ is the tangential shear of the source galaxies, $\Sigma_c$ is the same critical surface density defined in \autoref{eq:sigma-c}, and $w_i$ is a per-galaxy weight taken from the HSC shape catalog which characterizes the shape measurement error and intrinsic shape noise. We introduce random-source pairs because not all observed ellipticities are from gravitational lensing.  We use these random-source pairs to calibrate for these random alignments.  We specifically make use of the \textsc{python} package \texttt{dsigma}\footnote{https://github.com/johannesulf/dsigma} to carry out this measurement \citep{Lange22}.

\subsection{Populating N-body Simulations with Observables}
\label{subsec:sims}

We use the following formalism to create $\Delta \Sigma$ profiles that match the lensing signal measured in the data.  We populate halos in simulations with observables of varying scatter following \citet{Huang2022}. Briefly, we assume an analytic halo mass function and model our mass proxies as a log-linear relation with constant log-normal scatter, e.g.
\begin{equation*}
    \log \SO = \mathcal{N}(\alpha\log\M + \pi, \sigma_{\SO|\M}),
\end{equation*}
where $\alpha$ and $\pi$ are the slope and height of mass-observable relation, and $\sigma_{\SO|\M}$ is the scatter in observable $\SO$ at fixed mass $\M$.  \citet{Huang2022} demonstrated that there is a degeneracy between $\alpha$ and $\sigma_{\SO|\M}$:
\begin{align}
   \label{scattereq} \sigma_{\M|\SO} &= \left[ \left(\frac{\alpha}{\sigma_{\SO|\M}}\right)^2 + \beta \right]^{-1/2}\\
    \notag & \propto \frac{\sigma_{\SO|\M}}{\alpha},
\end{align}
where $\beta$ is the steepening slope of the HMF. We highlight that a given $\sigma_{\M|\SO}$ can be generated by different $\sigma_{\SO|\M}$ and $\alpha$. Our analysis does not differentiate between the two. We are not estimating the absolute value of the slope of the scaling relation for each mass proxy but rather quantifying the magnitude of the scatter in each observable, so we fix the slope of the scaling relation, $\alpha = 1$, and vary the scatter when populating the N-body simulations with observables.

In the simulation, we generate mock observables with varying scatter in observable at fixed mass, $\sigma_{\SO|\M}$, and use \eqref{scattereq} with $\alpha = 1$ to equate this to scatter in mass at fixed observable $\sigma_{\M|\SO}$. Hereafter we simply refer to this quantity as the scatter, $\sigma$, associated with each mass proxy.  We do this for a range of $\sigma$ values from 0 to 1.0 dex.  We use the virial mass distribution for each simulated observable to model the $\Delta \Sigma$ signal, resulting in a specific $\Delta \Sigma$ profile for each observable.

To find the ``best-fit'' model for a given measured $\Delta \Sigma$ signal, we minimize the quantity
\begin{equation}
\label{eq:chi2}
\chi^2 = (\Delta \Sigma_{\mathrm{Model}} - \Delta \Sigma_{\mathrm{Obs}})^\mathrm{T} \vb{C}^{-1} (\Delta \Sigma_{\mathrm{Model}} - \Delta \Sigma_{\mathrm{Obs}}),
\end{equation}
where $\Delta \Sigma_{\mathrm{Model}}$ is the predicted $\Delta \Sigma$ from our model, $\Delta \Sigma_{\mathrm{Obs}}$ is the measured $\Delta \Sigma$ in our data, and $\vb{C}$ is the covariance matrix, following the approach described in Appendix A of \citet{Huang2022}.  The covariance matrix was estimated using bootstrap resampling with 50,000 iterations. Jackknife resampling produced comparable results, but we favored bootstrap due to its ability to account for skewed distributions.  Since we are calculating the $\chi^2$ statistic on a finite grid of values of scatter, we interpolate the normalized cumulative distribution function (CDF) and take the 50th percentile as the best-fit model rather than taking the minimum $\chi^2$.  Specifically, we interpolate the CDF of our Likelihood, $\mathcal{L} = \exp \left( -\frac{\chi^2}{2} \right)$, over the grid of $\sigma$ values, we evaluate the CDF at the 14th, 50th, and 86th percentile to estimate the best-fit scatter and $1$-$\sigma$ uncertainties. 

Then, by using abundance matching to define equal number density bins in each observable (described in detail in \autoref{subsec:topn}), we directly compare the profiles on the same figure.  It is important to note that this Gaussian plus scatter-only model does not correctly model a lensing signal that contains systematics present in cluster selection, such as mis-centering and projection effects \citep{Sunayama2020}, but the overall amplitude of the model will still probe the scatter of the mass at fixed observable.

\section{Methods}\label{sec:methods}

We now describe our approach for categorizing clusters detected in $\srm$ and  $\smout$.  We list failure modes found when visually inspecting clusters detected in only one sample. Then, we describe the samples used for measuring weak lensing. Lastly, we detail our approach for fitting the $\lambda-\mout$ relation.

\subsection{Matching \redm clusters and HSC massive galaxies}
\label{subsec:matching}

\begin{figure*}
    \begin{center}
    \includegraphics[width=.95\textwidth]{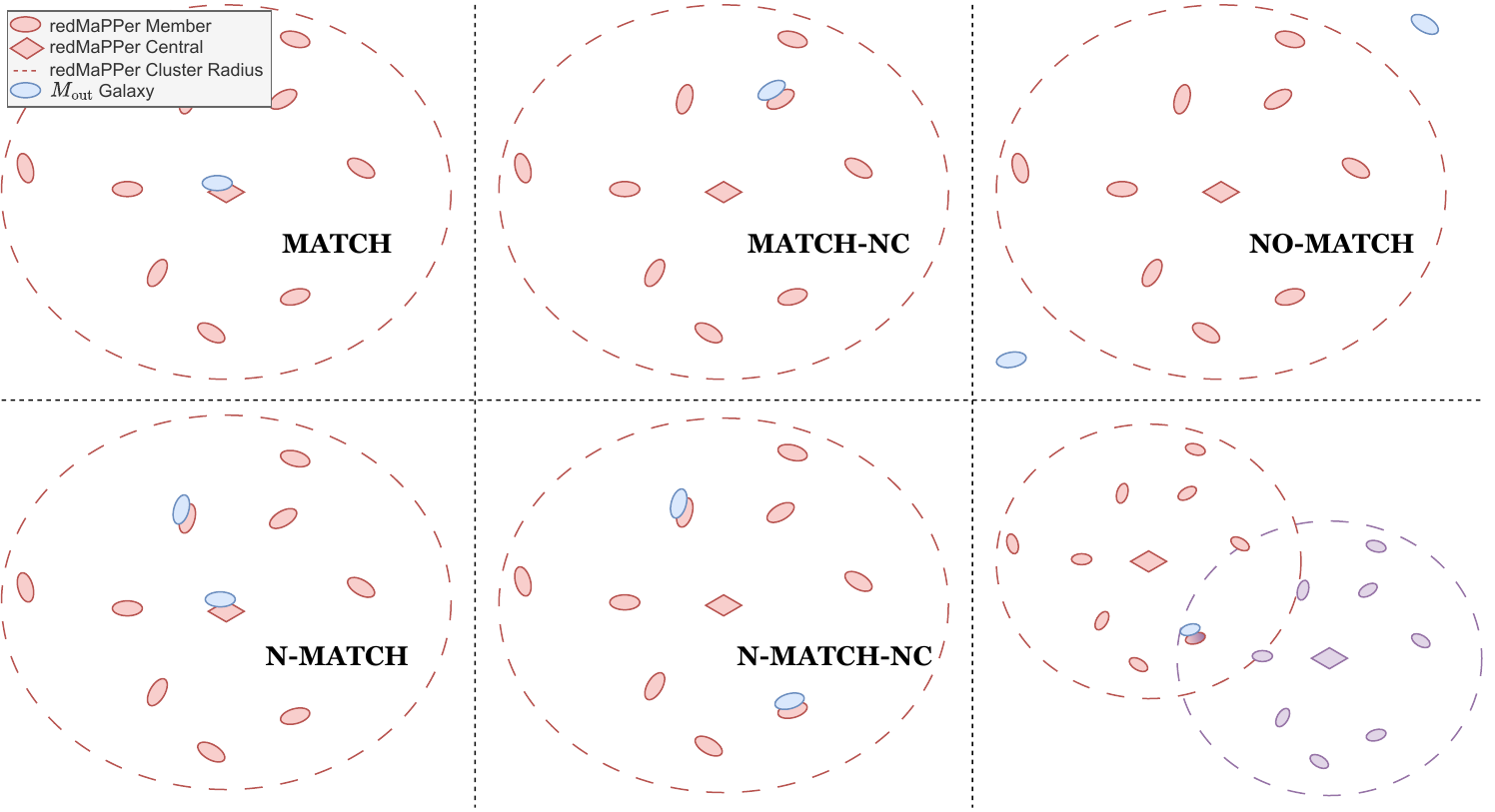}
\caption{Diagrams that illustrate how we categorize clusters detected in both $\smout$ and $\srm$ or only found in one sample.  \redm clusters are shown in red or purple, with dashed lines representing the cluster radius ($R_\lambda$), the \redm identified central galaxy as a diamond, and \redm members as red or purple ellipses.  The $\mout$ galaxies are shown in blue. The upper left shows the MATCH category, where a unique match between $\srm$ and $\smout$ and the matched $\mout$ galaxy is the CG.  The upper middle shows MATCH-NC, where there is again a unique match between $\smout$ and $\srm$, but the matched $\mout$ galaxy is not the CG.  The upper right shows NO-MATCH where a cluster in $\smout$ or $\srm$ has no match.  In the bottom left, we show N-MATCH, where a single \redm cluster has $N=2$ $\mout$ galaxies matched, and one of those is the CG.  Similarly, in the bottom middle, we show N-MATCH-NC where a single \redm cluster has $N=2$ $\mout$ galaxies matched, but neither are the CG.  Lastly, in the bottom right, we illustrate the possibility of two \redm clusters matching to a single $\mout$ galaxy; however, this does not occur within our sample.}
    \label{figure:match-categories}
    \end{center}
\end{figure*}

\subsubsection{Categorizing Matches between Catalogs}
\label{subsubsec:gal_cl_match}
We expect the $\smout$ to comprise roughly $90\%$ central galaxies and $10\%$ satellite galaxies (DeMartino et al. in preparation; \citealp{Huang2022, Saito16}), and we investigate this satellite contamination in this paper. We match galaxies from $\smout$ and $\srm$, allowing for a one arcsecond offset\footnote{We use the Hierarchical Triangular Mesh algorithm as implemented in \textsc{esutil} (https://github.com/esheldon/esutil) to perform this matching process.}. We assign the following categories to our matched galaxies and illustrate them in \autoref{figure:match-categories}.

\begin{itemize}
    \item \textbf{MATCH}: A \redm cluster matches to a single $\mout$ galaxy and that galaxy matches to the \redm central galaxy (CG).  
    \item \textbf{MATCH-NC}: A \redm cluster matches to a single $\mout$ galaxy, but that galaxy matches to a member galaxy that is not the CG. 
    \item \textbf{NO-MATCH}: All members of a given \redm cluster do not match to any $\mout$ galaxy, OR a given $\mout$ galaxy does not match to any \redm member.   
    \item \textbf{N-MATCH}: A \redm cluster has $N-\mout$ galaxies that match its members, and one of those $N$ members is the CG. In our diagram, we illustrate the case where $N=2$, but here, $N$ can take any number.  This value will be different depending on the match direction.  For example, we have 2 N-MATCH massive galaxies in the diagram shown but only 1 N-MATCH cluster.
    \item \textbf{N-MATCH-NC}: A \redm cluster has $N-\mout$ galaxies matched to its members, but none match to the CG.  Again, in our diagram, we illustrate the case when $N=2$.  As mentioned above, the value will also differ depending on the match direction.  For example, we have 2 N-MATCH-NC massive galaxies in the diagram but only 1 N-MATCH-NC cluster.
\end{itemize}
Because a single \redm member can belong to more than one \redm cluster, it is technically possible for a single $\mout$ galaxy to match to more than one \redm cluster, illustrated in the lower-right diagram in \autoref{figure:match-categories}.  We bring up this case to cover all possibilities involved in matching the samples but do not create a category because it does not occur in our sample. 

In our analysis, we assume that the MATCH category represents an agreement in cluster selection and $\sigma$. The MATCH-NC, N-MATCH, and N-MATCH-NC categories can be due to four scenarios: 
\begin{itemize}
\item  Correct centering of \redm with a satellite galaxy in $\smout$
\item  Mis-centering of \redm with the massive galaxy in $\smout$ as the true center
\item The massive galaxy in $\smout$ is along the line-of-sight and does not belong to the cluster.
\item The true cluster center was not detected by either $\smout$ or $\srm$. 
\end{itemize}
The number of $\smout$ galaxies in a single cluster correlates with the satellite fraction in $\smout$\footnote{specifically (N-MATCH-NC)-1 + (N-MATCH)-1} but is affected by both failure modes in the $\mout$ measurement and \redm membership accuracy.  Similarly, the number of \redm clusters with a matched central and satellite\footnote{specifically MATCH-NC + N-MATCH-NC)} correlates with \redm miscentering but can be affected by \redm centering accuracy. Therefore, the validity of \redm membership and centering assignments can affect our findings. In \autoref{sec:results}, we explore tests of \redm membership and centering determinations using spectroscopy and stacked lensing.

Because N-MATCH and N-MATCH-NC clusters can impact our $\Delta \Sigma$ measurement, \textbf{we remove percolated $\mout$ galaxies as described in \autoref{sec:runcat} in these categories when measuring the stacked $\Delta \Sigma$ profiles}. We consider the percolated galaxy to be a satellite. We include all other match categories as is.  

\subsubsection{Visual Inspection process}
\label{subsec:failmodes}
We visually inspect all clusters and count the incidence of the following failure modes that could cause the lack of a perfect MATCH categorization:
\begin{itemize}
    \item \textbf{CG-BLEND}: The central galaxy of a \redm cluster has a double core, which can cause the 1-D surface brightness profile extraction to fail \citep{Huang18}, and this galaxy is missing from the Huang et al. catalog.  
    \item \textbf{SAT-BLEND}: A \redm satellite of a cluster has a double core which can cause the same failure as in CG-BLEND, and no $\mout$ galaxy is found for that cluster.  This \redm satellite could be the true cluster center.
    \item \textbf{NF} (Not Found): An $\mout$ galaxy is within the $R_{\lambda}$ of a \redm cluster but this galaxy is not detected by the \redm algorithm.  We require the line of sight offset between galaxies to be $|z_{\textrm{HSC}} - z_{\lambda}| \leq 0.05$ to capture uncertainties in the $z_{\lambda}$ estimate \citep{Rykoff2014}. 
    \item \textbf{SS} (Super Spreader): A very bright object in the DES photometry can affect the local background of the cluster.  This effect can lead to spurious cluster member detections, increasing the richness of a \redm cluster or leading to the misidentification of false sources as the central.  This scenario may cause a false cluster detection in $\srm$ due to a boosted $\lambda$ measurement.
    \item \textbf{M} (Merge): A \redm cluster has no matched $\mout$ galaxy, but is within the vicinity of another \redm cluster that has a matched $\mout$ galaxy.  We do not require the clusters to be within each other's $R_\lambda$ but do require the same redshift criterion as for the NF failure mode.
    \item \textbf{CG-Z}: The $\zspec$ and $z_\lambda$ of a \redm cluster disagree. This disagreement may result in a cluster outside our redshift range included in $\srm$.
\end{itemize}

When identifying mergers, we note a considerable ambiguity in what constitutes a single cluster versus two independent merging systems.  Due to this ambiguity in the dynamical state of the virialized mass of a cluster, we consider many \textit{potential} mergers that may be two distinct clusters. 

\begin{figure*}
    \begin{center}
    \includegraphics[width=.33\textwidth]{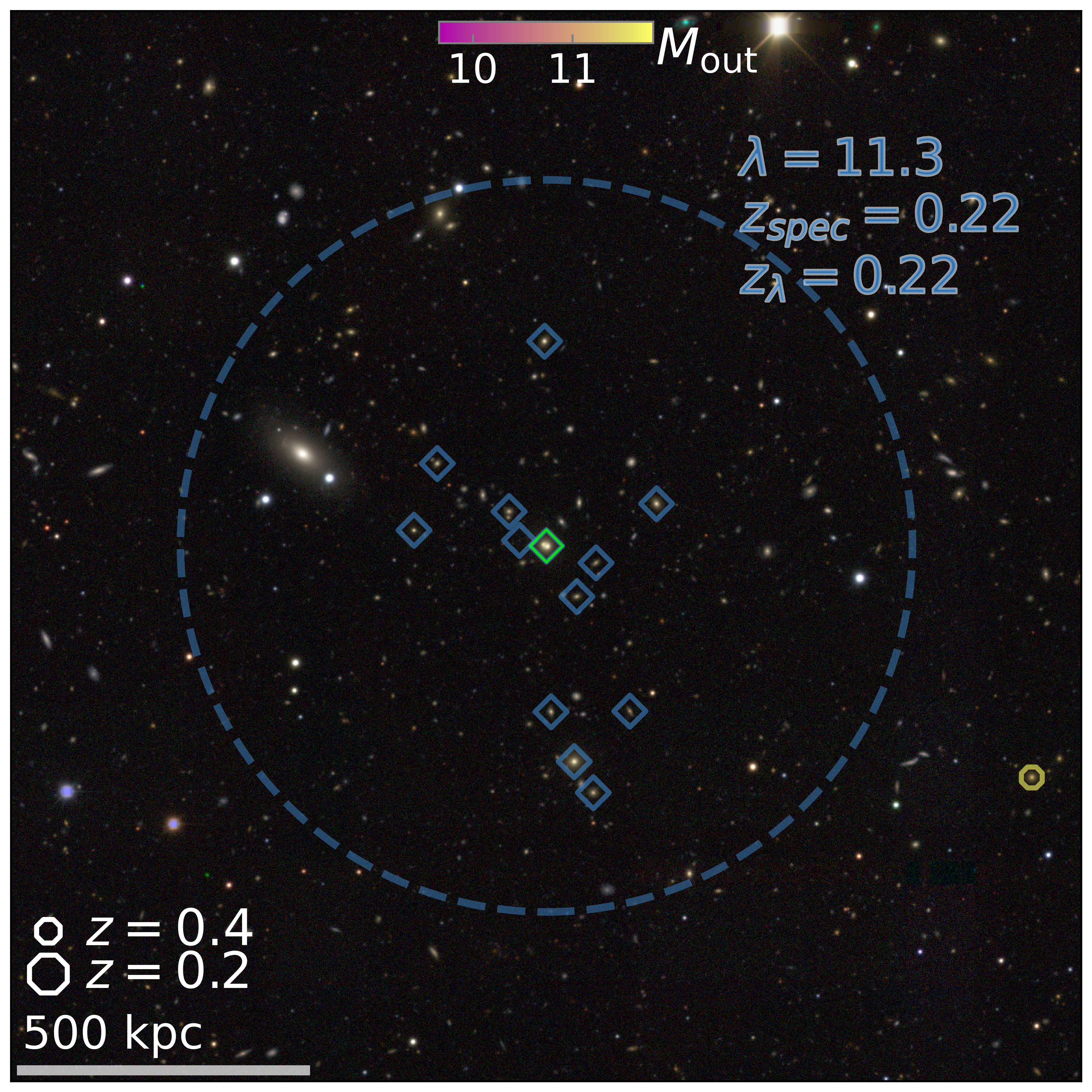}
    \includegraphics[width=.33\textwidth]{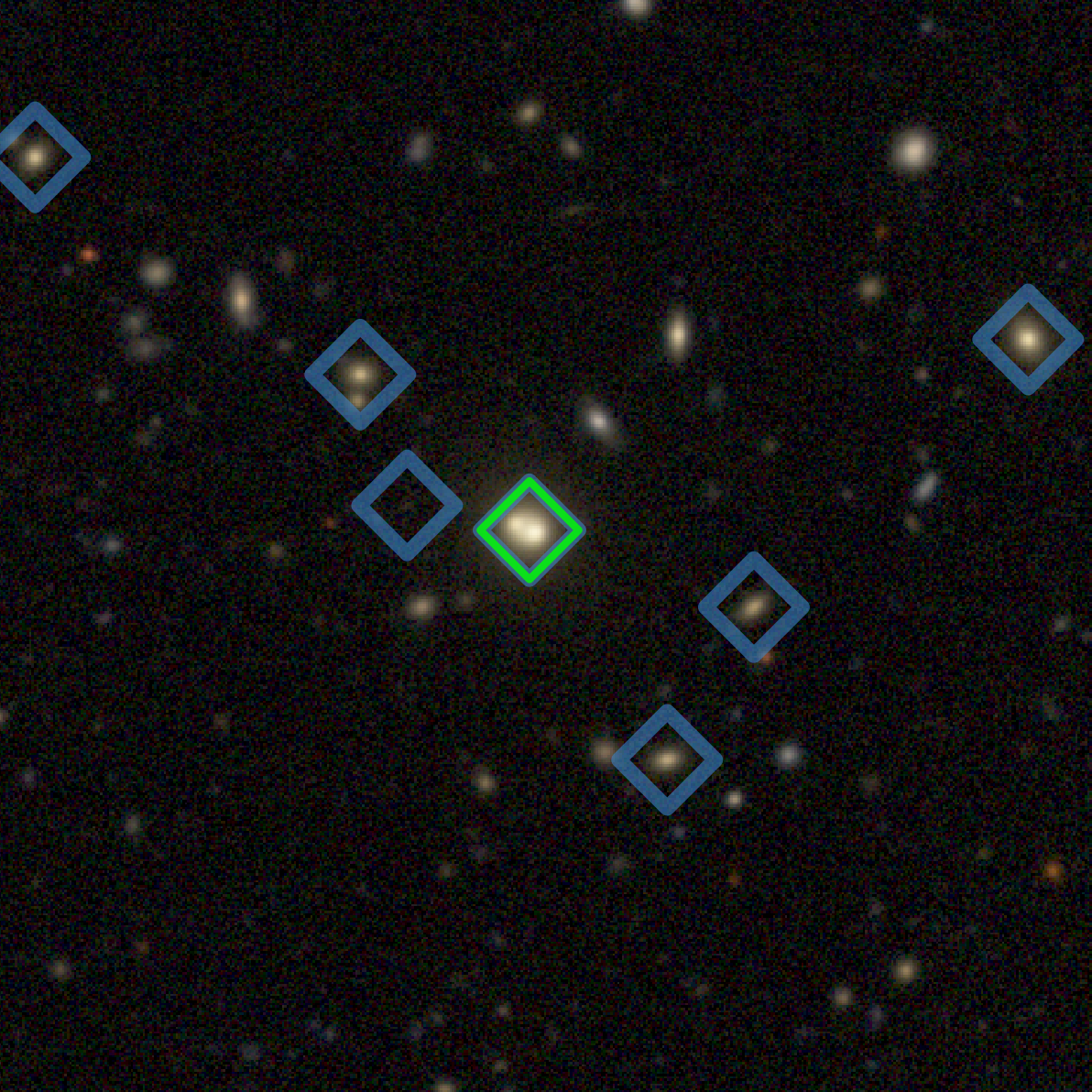}
    \includegraphics[width=.33\textwidth]{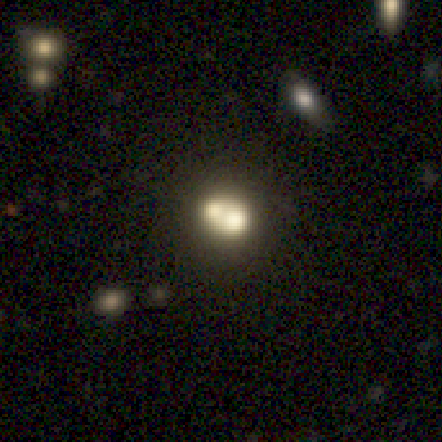}
    \caption{An example of the CG-BLEND failure mode.  On the left, we show a \redm cluster in blue, where the dotted line is $R_\lambda$, blue diamonds are cluster members, and the green diamond is the \redm CG.  A single $\mout$ galaxy in yellow in the bottom right is not matched to this \redm cluster (hence both are NO-MATCH).  The size of the markers shown in the lower left are indicative of each galaxy's redshift.  In the middle, we zoom into the central and see the CG-BLEND plus a potential SAT-BLEND to the upper left.  On the right, we zoom in further to the \redm CG and remove all markers.  It is clear that the \redm CG is a double-core galaxy, and we suspect that the double core causes the $\mout$ measurement to fail, resulting in this \redm cluster only being detected in $\srm$.}
    \label{fig:failure-modes}
    \end{center}
\end{figure*}

We show examples of our visual inspection process and the CG-BLEND failure mode in \autoref{fig:failure-modes}.  We visually inspect all NO-MATCH clusters in $\srm$ and $\smout$ and report the results in \autoref{subsec:no-match-vi}.  Then, we visually inspect all N-MATCH, N-MATCH-NC, and MATCH-NC clusters and report the findings in \autoref{subsec:n-match-vi}.

\subsection{$\Delta \Sigma$ as an estimator of $\sigma$}
\label{subsec:topn}

The following example can best illustrate the idea underlying our quantitative comparison of mass proxies. Suppose we compute the lensing signal of the top \textit{N} most massive halos in a sample. Now, given some mass proxy, $\SO$, with zero scatter (a perfect tracer), we can rank order all halos selected by $\SO$ and again compute the lensing signal of the top \textit{N} selected halos. Because the tracer has no scatter, the measured lensing signals will be identical to picking the top \textit{N} most massive halos.  As we add scatter to the mass proxy, $\SO$, less massive halos will scatter into the selection, and the lensing signal will become shallower.  Thus, the amplitudes of the best-fit $\Delta \Sigma$ models can be used to make relative comparisons between mass proxies. 

We define bins in $\mout$ and $\lambda$ that correspond to the same number density of objects.  To do so, we find the number density of halos in an N-body simulation. Then, we take the assumed mass function to find the halo mass cut which results in the same number of halos.  Using this approach we connect a given number density of halos to a specific mass range. Then, we use the fact that the HSC sample of massive galaxies is nearly complete for $0.19 < z < 0.51$ and $\log_{10}({M_{\star,100\kpc}/M_\odot}) \geq 11.6$ \citep{Huang2022} to define bins in $\mout$ with equivalent number density in $\mout$.  Lastly, we define bins in $\lambda$ such that the number densities in $\lambda$ are the same.  The bins for $M_{\textrm{vir}}$, $\lambda$, and $\mout$ are given in \autoref{table:bins}. 

\begin{table*}
    \centering

    \begin{tabular}{@{}ccccc}
    \hline
    Property & Bin 1 & Bin 2 & Bin 3 & Bin 4\\
    \hline
        $M_{\textrm{vir}}$   &$[14.66, 15.55]$   & $[14.38, 14.66)$    & $[14.08, 14.38)$ & $[13.86, 14.08)$\\
        $\mout$     &$[11.20,11.60] $   & $[11.0, 11.20)$    & $[10.8,11.0)$ & $[10.63,10.8)$\\
        $\lambda$   &$[40,120] $        & $[20,40)$    & $[10,20)$    & $[6,10)$\\
    \hline
    \end{tabular}
    
     \caption{Equal number density bins in $M_{\textrm{vir}}$, $\mout$, and $\lambda$ used in our analysis.  We obtain the $M_{\textrm{vir}}$ bins by counting halos in MDPL2 and SMDPL simulations and define the corresponding $\mout$ bins using the mass complete $\smout$ sample.  We then define our $\lambda$ bins to match the number density of these $\mout$ bins.}
     
    \label{table:bins}
\end{table*}

\subsection{Samples}\label{subsec:samples}

Using the match categories and number density bins defined above we create three different samples of clusters.  We describe those samples below.  These are the clusters and galaxies that we make weak gravitational lensing measurements on.

\subsubsection{Number Density Sample}\label{subsec:topn-sample}
This sample consists of all match categories between $\smout$ and $\srm$, removing satellite $\mout$ galaxies.  Then, we bin these matches according to the number density bins defined in \autoref{table:bins}, and subdivide each number density bin by the other mass proxy. By subdividing each bin by mass proxy, we can assess whether there is additional halo mass information in either mass proxy and compare the magnitude of $\sigma$ for each proxy.    

As an example for one $\lambda$ bin, we take all \redm clusters and then subdivide those clusters according to the $\mout$.  Some clusters in this $\lambda$ bin will scatter high in $\mout$, and some will scatter low.  We then measure the $\Delta \Sigma$ amplitude, obtain a best-fit model for clusters that scatter high and low (or remain consistent), and compare it to the total $\lambda$ sample.  A difference or consistency in amplitude tells us about each proxy's information about the underlying mass and the accuracy of each proxy measurement.  We repeat this process for all $\lambda$ bins and all $\mout$ bins.

\subsubsection{Sliding Conditional Percentile Sample and Proxy Dependent Scatter}
Next, we define the Sliding Conditional Percentile (SCP) sample to understand $\sigma$ as a function of halo mass.  We take all match categories between $\srm$ and $\smout$ and remove satellite $\mout$ galaxies.  We use the \textit{sliding conditional percentile} weighting, as implemented in \texttt{halotools v0.8}\footnote{https://github.com/astropy/halotools} \citep{Hearin17}, which estimates the cumulative distribution function $P(<y|x)$  to create bins in $\lambda$ at equal $\mout$, and bins in $\mout$ at equal $\lambda$.  We weigh the value of each of the secondary halo properties by selected primary property. These weights are uniformly distributed between 0 and 1. We rank the weights and bin them into percentiles. This method guarantees that all bins have the same primary property distribution. For a given bin in $\lambda$ at fixed $\mout$, we measure $\Delta \Sigma$, obtain a best-fit model, and compare the best-fit model between all $\lambda$ bins.  We then do the reverse for bins in $\mout$ at fixed $\lambda$.  When we find the best-fit $\Delta \Sigma$ model for the observed signal, we use number density bin two, which corresponds to the fixed proxy values of $20 > \lambda \geq 10$ and $11 > \mout \geq 10.8$. We use this test to see the halo mass dependence of $\sigma$ for both $\lambda$ and $\mout$ while ensuring that any differences in scatter do not originate from skewed binning.  We show the mean values in $\lambda$ bins at fixed $\mout$ and the mean values in $\mout$ bins at fixed $\lambda$ in the top and bottom rows of \autoref{table:scp_bins}.

\begin{table}
    \centering

    \begin{tabular}{@{}ccccc}
    \hline
     & Bin 1 & Bin 2 & Bin 3 & Bin 4\\
    \hline
    $\lambda$ & $7.1 \pm 0.13$ & $10 \pm 0.2$ & $15 \pm 0.4$ & $27 \pm 1.5$\\
    $\mout$ & $10.86 \pm 0.022$ & $10.86 \pm 0.021$ & $10.86 \pm 0.022$ & $10.88 \pm 0.026$\\
    \hline
    \hline
    $\lambda$ & $14 \pm 1.2$ & $16 \pm 1.3$ & $14 \pm 1.0$ & $15 \pm 1.5$ \\
    $\mout$ & $10.68 \pm 0.004$ & $10.78 \pm 0.009$ & $10.91 \pm 0.010$ & $11.11 \pm 0.017$\\
    \hline
    \end{tabular}
    
     \caption{The mean value of $\lambda$ and $\mout$ for each bin in the sliding conditional percentile sample, where errors reported are the standard error of the mean. The small error demonstrates the mean is a good representative of the sample.  The top row shows bins in $\lambda$ at approximately equal $\mout$, and the bottom shows bins in $\mout$ at approximately equal $\lambda$.  We define these bin edges using sliding conditional percentile weighting to prevent issues from skewed bins.}
    \label{table:scp_bins}
\end{table}

\subsubsection{Surjective and single-detection Samples}
To assess cluster selection, we take all match categories and remove satellite $\mout$ galaxies. This results in \redm clusters that have a matched $\mout$ galaxy (i.e., detected in both $\srm$ and $\smout$) and clusters that do not (i.e., not detected in $\srm$). Clusters that are detected in both catalogs are defined as the surjective sample, and clusters with a single-detection in either $\srm$ or $\smout$ are defined as the single-detection sample.  We obtain a surjective and single-detection sample for each number density bin and each mass proxy.

We measure the $\Delta \Sigma$ signal and find the best-fit model for the surjective and single-detection clusters for a given $\lambda$ ($\mout$) bin.  We compare the counts and $\sigma$ for each of these samples.  To ensure any differences in amplitude are not from skewed bins, we randomly draw from the more numerous group (e.g., either matched or missing) using a Metropolis-Hastings algorithm \citep{Metropolis53, Hastings70} to match the distribution of the smaller sample. Then we re-measure the lensing signal, fit a best-fit model, and compare the resulting $\Delta \Sigma$ and $\sigma$.  A systematic difference between the $\Delta \Sigma$ signal for matched and missing groups may indicate a lack of purity in either $\srm$ and $\smout$, as a less pure sample would have a shallower amplitude of the best-fit model.

\section{Results}\label{sec:results}

\subsection{Characterizing the Two Cluster Samples}\label{subsec:results_match}

We quantify the incidence of failure modes in each cluster sample and visually inspect clusters identified by each mass proxy.  Then, we asses the cluster and satellite detection in each sample.

\subsubsection{Failure Modes in NO-MATCH Categories}\label{subsec:no-match-vi}

We show the number of occurrences of each failure mode for each sample in \autoref{table:no-match-stats}. We find 363 clusters detected in $\srm$ with no detection in $\smout$.  Many of these are classified as CG-BLEND or SAT-BLEND, with counts of 102 and 25, respectively, making up roughly one-third of all of the unmatched clusters.  A double-core can cause a galaxy's extracted 1-D surface brightness profile to show an upturn, %when a bright companion is nearby.  
and the $\mout$ measurement can fail in this case or for systems with a bright companion if the companion galaxy cannot be fully masked.  This is a known failure mode of the photometry pipeline used to estimate the stellar masses within $\smout$; an additional failure mode is the presence of bright stars or foreground galaxies. In total the measurements failed for $\sim 10 \%$ of galaxies in the original massive galaxy sample \citep{Huang18}.  This percentage is less than the percentage of CG-BLEND and SAT-BLEND cases visually identified, but almost certainly is one source of these unmatched clusters. We find 16 \redm clusters in the not found (NF) category.  It is unclear why $\mout$ selection does not detect a cluster in these cases. 

327 clusters are detected in $\smout$ and not detected in $\srm$.  Unlike $\srm$, we do not find a dominant failure mode.  The highest occurrence of a failure mode is 21 instances of CG-BLEND, where the $\mout$ galaxy is in a crowded field, which could result in an overestimated $\mout$ value.  We find 19 NF clusters, where it is unclear why \redm does not consider the $\mout$ galaxy a member. 

These results show that double-cores are the dominant failure mode that causes clusters only to be detected in $\srm$ or $\smout$.   These blends can cause issues for both cluster selections.  For \redm, the total luminosity from multiple cores is considered a single source.  This combination leads to overestimating the luminosity and potentially incorrect CG for the cluster.  this scenario increases the number of MATCH-NC and NO-MATCH clusters. For $\mout$, N-MATCH-NC, MATCH-NC, N-MATCH, and NO-MATCH categories can also be affected by blending and crowded fields that cause the $\mout$ measurement to fail.  First, the number of NO-MATCH clusters in $\srm$ would increase, which again we see a higher number in $\srm$ compared to $\smout$.  The number of N-MATCH-NC and N-MATCH categories would increase due to a massive galaxy SAT-BLEND that $\mout$ fails to measure.

The remaining failure modes rarely occur in $\srm$ or $\smout$.  In $\srm$ there are two occurrences of the merge (M) failure mode. However, as discussed previously, the definition of a merging system is nebulous, and we remained conservative with our judgment.  Including X-ray data in the visual inspection could help identify mergers more concretely.  We find four super-spreader (SS) failure modes.  These four clusters have a low richness before excluding the false cluster member detections.  We believe these clusters are too low mass to be valid candidates for matching in $\srm$.  Similarly, we find five instances of the CG-Z failure mode.  In these five cases, the CG was incorrectly chosen by \redm and is at a much different redshift than $z_\lambda$.  Upon visual inspection, it is clear $z_\lambda$ is the correct redshift for all five CG-Z clusters and outside of our initial redshift cut.

\subsubsection{Visual inspection of N-MATCH-NC, N-MATCH, and MATCH-NC\label{subsec:n-match-vi}}

We begin with clusters in $\srm$.  There are three N-MATCH-NC clusters, with N=2 for two of the clusters and N=3 for the last.   For two of these, we see a CG-BLEND failure mode, which could explain the lack of $\mout$ detection at the central.  For the third, we find a SAT-BLEND, which could affect \redm centering if that galaxy is the true \redm central.  Twenty-five clusters fall into the MATCH-NC category.  The CG-BLEND failure mode is present in 8 of these clusters.  This failure mode could be causing the $\mout$ measurement to fail for the CG or \redm to be mis-centered.  

There are 35 N-MATCH clusters, with N=3 for two of these and N=2 for the remaining 33. For 22 N-MATCH clusters, the \redm central is the more massive galaxy, which suggests the other matched $\mout$ galaxies are satellites.  The other 13 clusters could be cases of \redm mis-centering.

\begin{table*}
    \centering
    \begin{tabular}{lccccccc}
    \hline
    & NO-MATCH & CG-BLEND & SAT-BLEND & NF & M & SS & CG-Z \\
    \hline
    $\srm$ & 363 & 102 & 25 & 16 & 2 & 4 & 5 \\
    $\smout$ & 327 & 21 & -- & 19 & -- & -- & --  \\
    \hline
    \end{tabular}
    \caption{Incidence of non-matches in $\srm$ or $\smout$.  NO-MATCH is the count of clusters only detected in one sample, CG-BLEND is the number of central galaxies that have a visible blend, SAT-BLEND is the number of satellite galaxies that have a visible blend, NF is the not found failure mode, M is the merge failure mode, SS is the super-spreader failure mode, and CG-Z is the \redm central galaxy redshift failure mode.  The dominant failure mode for both samples is the occurrence of blending, which could cause either \redm mis-centering or a failure to obtain an $\mout$ measurement.}
    \label{table:no-match-stats}
\end{table*}

\subsubsection{Cluster Detection}

We summarize the detected fraction and cluster categorization in \autoref{table:category_stats}.  $40 \pm 2\%$ of all \redm clusters in $\srm$ are detected in $\smout$.  We impose a $\lambda \geq 20$ cut and see the detected fraction sharply rise to $71 \pm 6 \%$.  This cut is typically used in cluster cosmology \citep{DESY1-clusters} because $\lambda < 20$ \redm clusters start to decrease in purity \citep{Farahi16}.  For clusters detected in $\smout$,  the detected fraction is slightly higher at $46 \pm 2\%$, and when making our $\lambda \geq 20$ equivalent to a $\mout \geq 11$ cut, the detected fraction is again slightly higher at $78\pm 4\%$. However, both match rates are consistent within the standard error, which suggests they have a consistent purity.   From this, we can see that most clusters only detected by one mass proxy are at low halo mass and specifically at halo masses lower than what is included in cluster cosmology.

\begin{table*}
    \centering
    \renewcommand{\arraystretch}{1.5}
    \begin{tabular}{lcccccc|cc}
    \hline
    & Total & Matched & MATCH & MATCH-NC & N-MATCH & N-MATCH-NC & Duplicates & RM Satellites \\
    \hline
    $\srm$ &	603&	$40 \pm 2 \%$ ($240$)&	$74 \pm 3 \%$ ($177$)&	$10 \pm 2 \%$ ($25$)&	$15 \pm 2 \%$ ($35$)&	$1 \pm 1 \%$ ($3$)&$16 \pm 2 \%$ ($38$)&	$26 \pm 3 \%$ ($63$)\\
    $\lambda \geq 20$&	66&	$71 \pm 6 \%$ ($47$)&	$47 \pm 7 \%$ ($22$)&	$21 \pm 6 \%$ ($10$)&	$26 \pm 6 \%$ ($12$)&	$6 \pm 4 \%$ ($3$)&$32 \pm 7 \%$ ($15$)&	$53 \pm 7 \%$ ($25$)\\
    \hline
    $\smout$ &	608&	$46 \pm 2 \%$ ($281$)&	$63 \pm 3 \%$ ($177$)&	$9 \pm 2 \%$ ($25$)&	$26 \pm 3 \%$ ($72$)&	$2 \pm 1 \%$ ($7$) &$16 \pm 2 \%$ ($44$)&	$25 \pm 3 \%$ ($69$)\\
    $\mout \geq 11$&	85&	$78 \pm 4 \%$ ($66$)&	$86 \pm 4 \%$ ($57$)&	$8 \pm 3 \%$ ($5$)&	$6 \pm 3 \%$ ($4$)&	$0 \pm 0 \%$ ($0$)&$3 \pm 2 \%$ ($2$)&	$10 \pm 4 \%$ ($7$)\\
    \hline
    \end{tabular}
    \caption{The number of clusters detected in each catalog and the results of our match categorization for $\srm$ and $\smout$.  The top two rows show the results when matching $\srm$ to $\smout$, whereas the bottom two rows show the same results when matching in the opposite direction.  The MATCH, MATCH-NC, N-MATCH, and N-MATCH-NC categories are described in \autoref{subsubsec:gal_cl_match}.  We combine the incidence of N-MATCH and N-MATCH-NC categories into the Duplicates column and the N-MATCH, N-MATCH-NC, and MATCH-NC categories into the RM (\redm) Satellites column.  The incidence of duplicate matches contains information about satellite galaxies in $\smout$, and the incidence of \redm Satellites contains information about mis-centering in $\srm$. We first report the numbers for the entire sample, then impose a $\lambda \geq 20$ cut and equivalent $\mout \geq 11$ cut to maximize purity in each cluster sample.}
    \label{table:category_stats}
    \renewcommand{\arraystretch}{1}
\end{table*}

\subsubsection{Duplicate Matches and \redm Satellites}

In \autoref{table:category_stats} we report the number of clusters and massive galaxies in each of the N-MATCH, MATCH-NC, and N-MATCH-NC categories which are cases where the two cluster proxies may disagree, and add two derived columns: Duplicates and RM (\redm) Satellites.  The Duplicates count more than one $\mout$ galaxy matches to a \redm cluster, and the RM Satellites count the number of times a \redm satellite matches to an $\mout$ galaxy.  Percolating on $\mout$ resulted in 37 satellite galaxies. We note that this is slightly less than the total number of duplicates found because of the N-MATCH and N-MATCH-NC categories.

We see that at higher $\lambda$, a \redm cluster is much more likely to contain more than one $\mout$ galaxy.  Conversely, at high $\mout$, an $\mout$ galaxy rarely shares a \redm cluster with another massive galaxy.  These trends follow our expectations for the number densities and halo occupation of massive galaxies, in that massive galaxies become exceedingly rare at high mass while satellite galaxies become more common at high mass.  In \autoref{subsec:disc-satellites}, we speculate on the satellite fraction and mis-centered fraction of $\smout$ and $\srm$, respectively.  

\subsection{Lensing}
We now report the results when measuring weak gravitational lensing on clusters in the MATCH, MATCH-NC, N-MATCH, and N-MATCH-NC categories after removing duplicate matches to N-MATCH and N-MATCH-NC clusters as described in \autoref{subsec:matching}.  We specifically measure $\Delta \Sigma$ on these clusters in the different subsamples described in \autoref{subsec:samples}.

\subsubsection{Comparison of surjective and single-detection samples} \label{subsec:mamiss}

\begin{figure*}
    \begin{center}        
        \includegraphics[width=.49\textwidth]{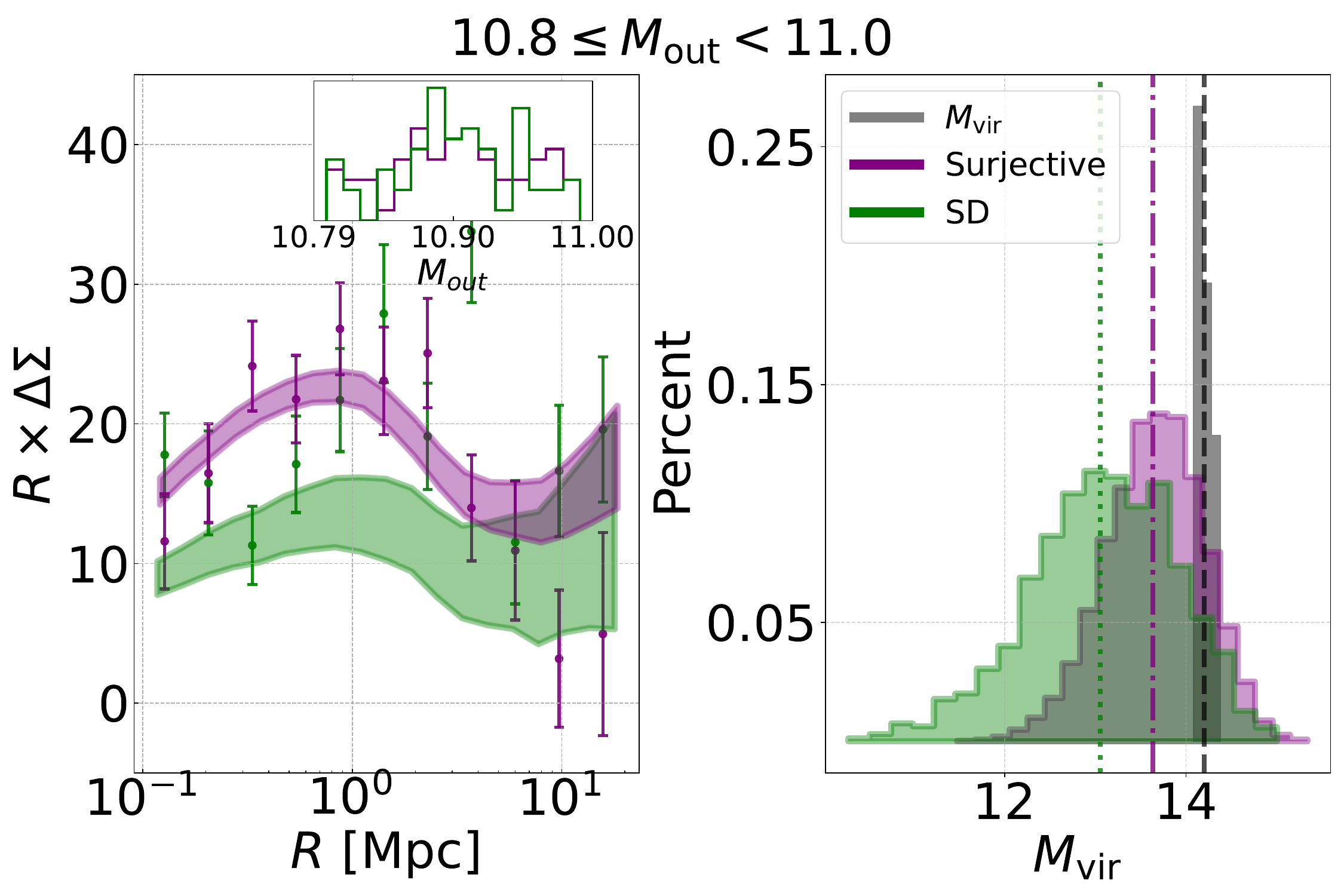}
        \includegraphics[width=.49\textwidth]{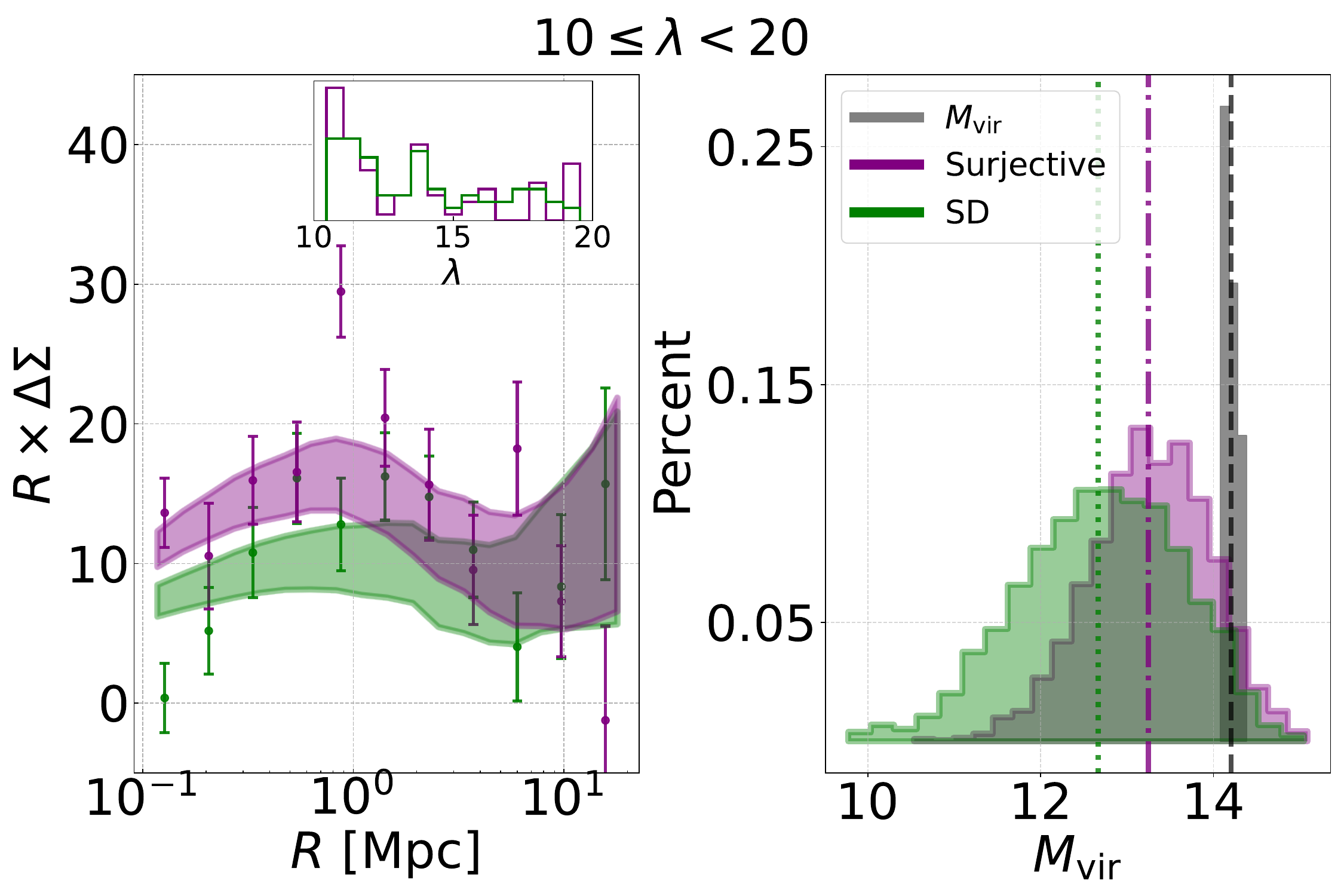}
        \includegraphics[width=.49\textwidth]{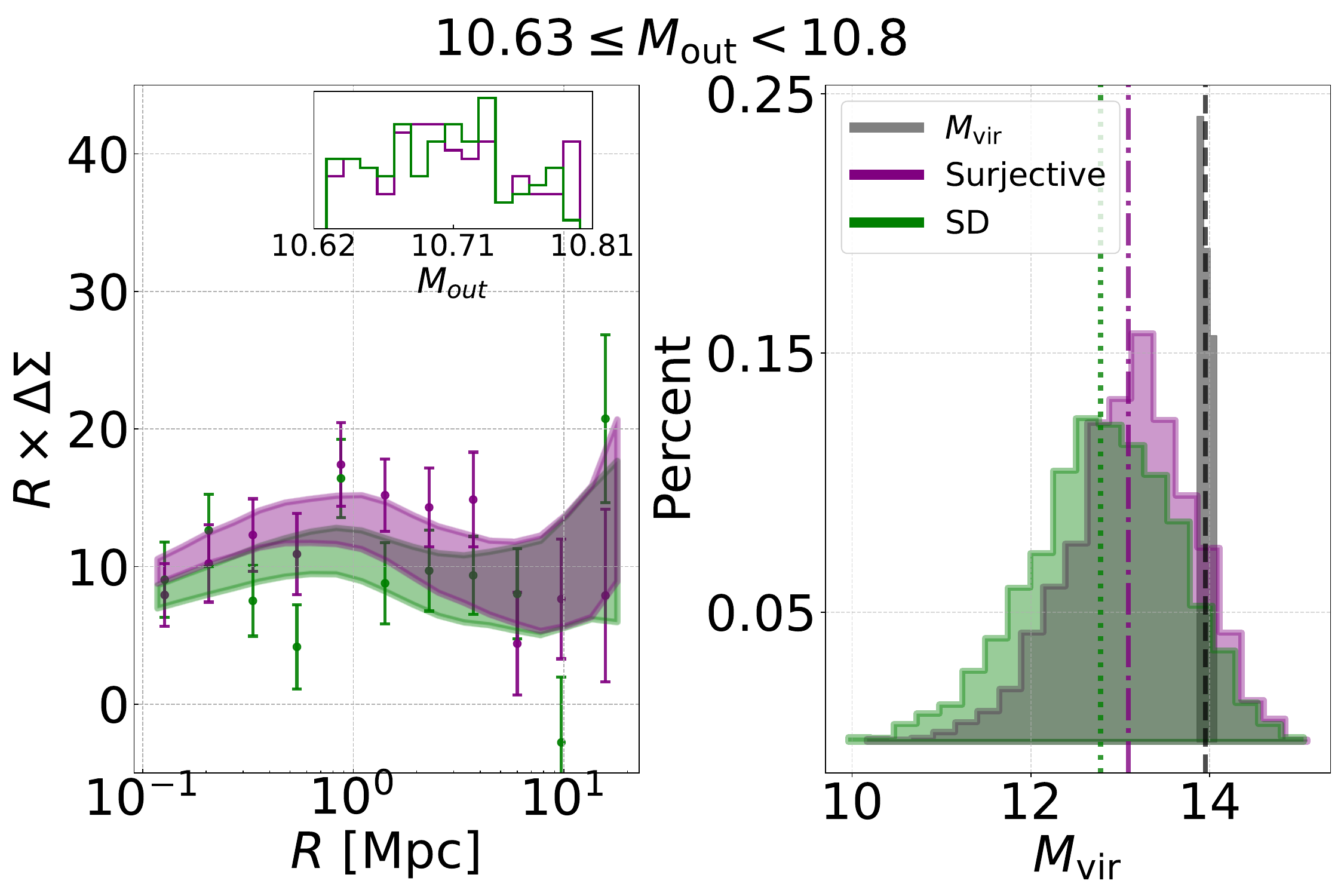}
        \includegraphics[width=.49\textwidth]{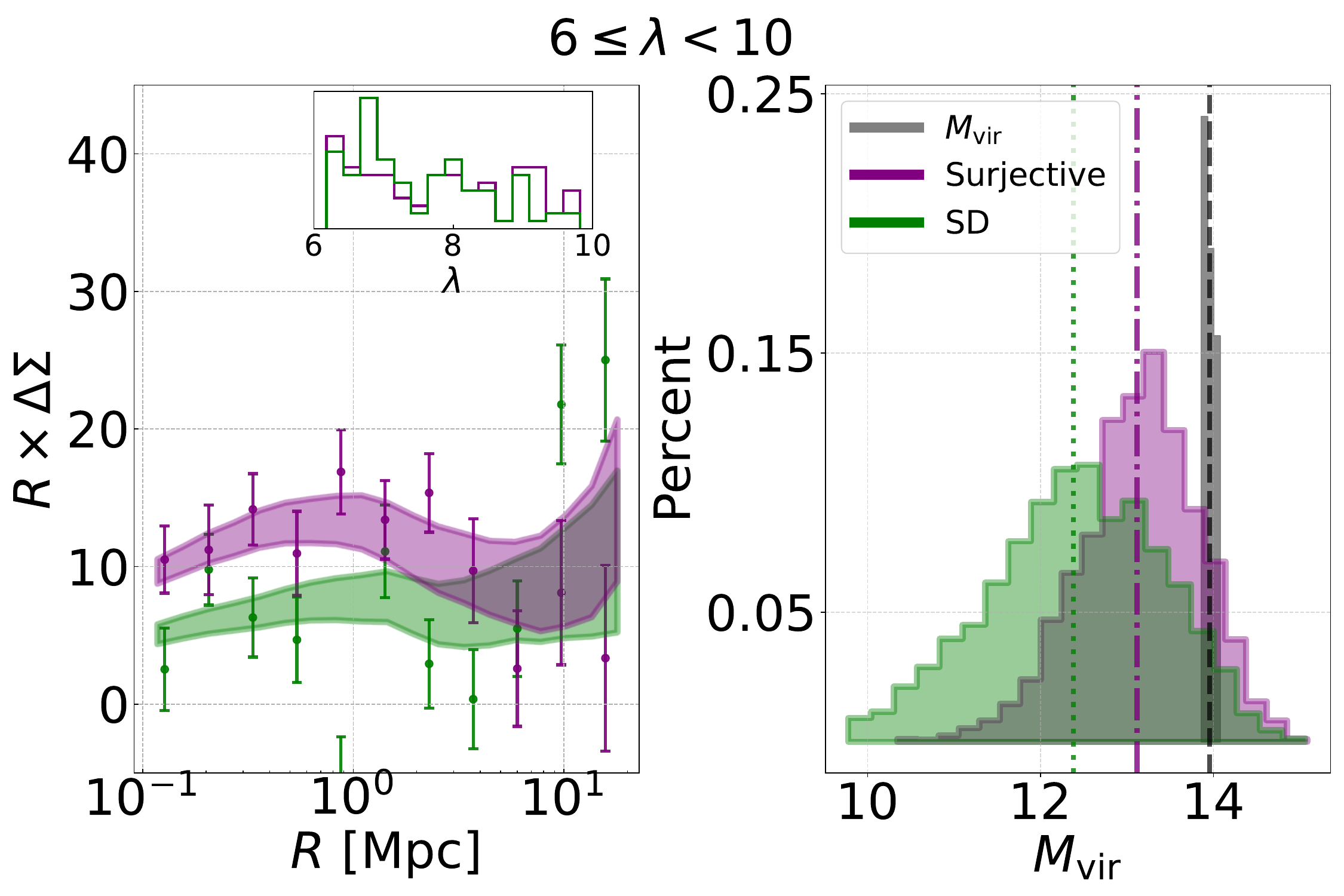}
        
        \caption{Comparison of the amplitude and scatter of surjective clusters to single-detection clusters in $\srm$ or $\smout$.  In the top row, we show $10.8 \leq \mout < 11.0$ bin on the left and $10 \leq \lambda < 20$ on the right.  This figure shows the bump in the $\lambda$ selected sample.  In the bottom row, we show $10.63 \leq \mout < 10.8$ on the left and $6 \leq \lambda < 10$ on the right. The left panel compares $\Delta \Sigma$ profile amplitudes for the surjective and single-detection samples, and the inset panel shows the distribution of $\mout$ or $\lambda$ within those samples.  The right panel shows the $M_{\textrm{vir}}$ distribution within simulations that generate the model, with the distribution for a perfect tracer shown in grey.  We show the surjective sample in purple and the single-detection sample in green.}
        \label{fig:mnm-bin}
    \end{center}
\end{figure*}

The lensing signals and associated $\sigma$ for the surjective and single-detection  samples are shown in \autoref{fig:mnm-bin}.  Both rows show a significant difference in $\Delta \Sigma$ amplitude between surjective and single-detection clusters for $\smout$ and $\srm$ for number density bins 3 (top) and 4 (bottom).  This difference in amplitude corresponds to differences in scatter.  For single-detection clusters in $\lambda$, $\mout$ bins 2 and 3, we measure a scatter of $(0.92, 0.80)$ and $(0.97, 0.81)$ respectively.  This is in contrast to the scatter measured for the surjective sample for $\lambda$, $\mout$ in bins 2 and 3 at $(0.69, 0.53)$ and $(0.67, 0.69)$ respectively. From this we see that clusters only detected in one catalog tend to be extremely poor tracers of halo mass with large scatter.

We see a clear bump in the $\Delta \Sigma$ profile for $\srm$ clusters around $R\sim 1\Mpc$.  We show a similar difference between surjective and single-detection samples in the lower panel for less rich clusters and less massive galaxies but to a lesser extent.  We use a Metropolis-Hastings algorithm to randomly draw from the larger sample to ensure the mass distribution within each sample is not biasing the lensing measurements.  We perform this random draw and re-measurement of the lensing signal multiple times to ensure the result is consistent.  We do not see any noticeable change when doing so and, therefore, show a single iteration in the figure.

We find that the number of single-detection \redm clusters is larger than the number of single-detection $\mout$ galaxies in the lowest richness bin, meaning more clusters are only detected in $\srm$ compared to $\smout$.  \redm selects more of this single-detection sample at lower $\lambda$ than the corresponding $\mout$ selection.  From number density bins two, three, and four, the match rate for $\lambda$ is $71 \pm 6 \%$, $54 \pm 4 \%$, and $26 \pm 2\%$.  Whereas for $\mout$ we have $74 \pm 5 \%$, $48 \pm 4\%$, and $37 \pm 3 \%$.   This could be driven either by a decrease in purity or completeness, but our lensing signals indicate this is most likely purity.  Thus,  we are seeing the purity fall off slower for $\mout$ than for $\lambda$, which could enable probing to lower halo mass with $\mout$ selected clusters.

\subsubsection{Scatter dependence on $\lambda$, $\mout$}
\label{subsec:chi-sq}
\begin{figure*}
    \begin{center}
    \includegraphics[width=.95\textwidth]{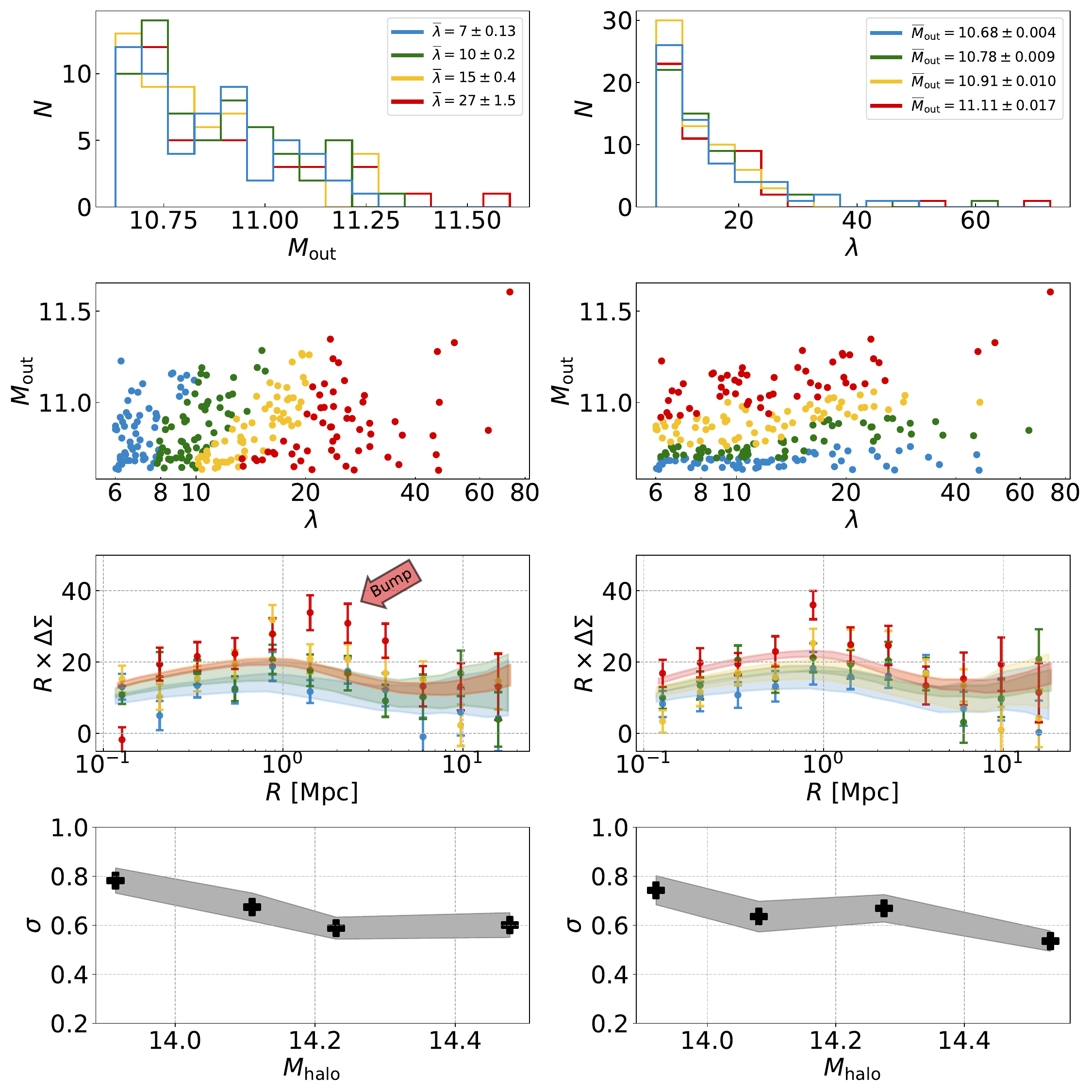}
    \caption{Top: The distribution of $\mout$ ($\lambda$) within each bin.  Second row: the distribution of cluster matches when binning the matches in $\lambda$ at fixed $\mout$ (left) and in $\mout$ at fixed $\lambda$ (right).  Third row: The best-fit $\Delta \Sigma$ models for the clusters within each bin. Bottom row: the best-fit $\sigma$ for each bin. Each color corresponds to a single SCP bin as defined in \autoref{table:scp_bins}, where blue, green, yellow, and red are bins one, two, three, and four respectively. We see consistent $\Delta \Sigma$ profiles for all bins except for the highest $\lambda$ and highest $\mout$ bins.  We highlight the presence of a ``bump'' in the $\lambda$ bins that appear at high richness with a red arrow and see the scatter of each mass proxy with the virial mass is very similar.
    }
    \label{fig:scp_bin_comp}
    \end{center}
\end{figure*}

Using the sliding conditional percentile sample, we compare our $\Delta \Sigma$ profile amplitudes. We plot our results in \autoref{fig:scp_bin_comp}.  From this figure, we can see two main results.  First, $\sigma$ for $\lambda$ and $\mout$ are consistent across the sampled range of halo masses.   We also highlight this consistency does not originate from skewed bins.  Secondly, we see the presence of a large ``bump'' around 1-3 $\Mpc$ for the $\overline{\lambda} = 24$ bin, highlighted with an annotated arrow.  The impact of this bump on $\Delta \Sigma$ modeling can be seen in the $\chi^2$ value for the best-fit model.  For the $\lambda$ selected model, the highest richness bin has a $\chi^2=3.56$, while the highest $\mout$ bin has a $\chi^2 = 1.98$. Here the $\chi^2$ is determined as described in Section \ref{subsec:sims} and equation \ref{eq:chi2}.    

Given that our best-fit models are log-linear relationships with constant scatter, this bump may originate from systematic biases that would need more sophisticated modeling to capture.  The increase in signal at these points pulls the best-fit amplitude of the $\Delta \Sigma$ model higher.  Therefore, these features may affect the inferred scatter of the $\lambda$ sample.

\subsubsection{Number density sample}\label{subsec:result-number-density}
Using the amplitude for the best-fit $\Delta \Sigma$ model as an estimator for $\sigma$, we compare the scatter of $\lambda$ to $\mout$ in the number density sample as described in \autoref{subsec:topn-sample}.  We do not have the required statistics for the lowest number density bin (highest mass proxy) to make meaningful lensing measurements.  For this reason, we do not include the bin containing $40 \leq \lambda \leq 120$ and $11.2 \leq \mout \leq 11.6$ in our results.  

\begin{figure*}
    \begin{center}
    \includegraphics[width=.49\textwidth]{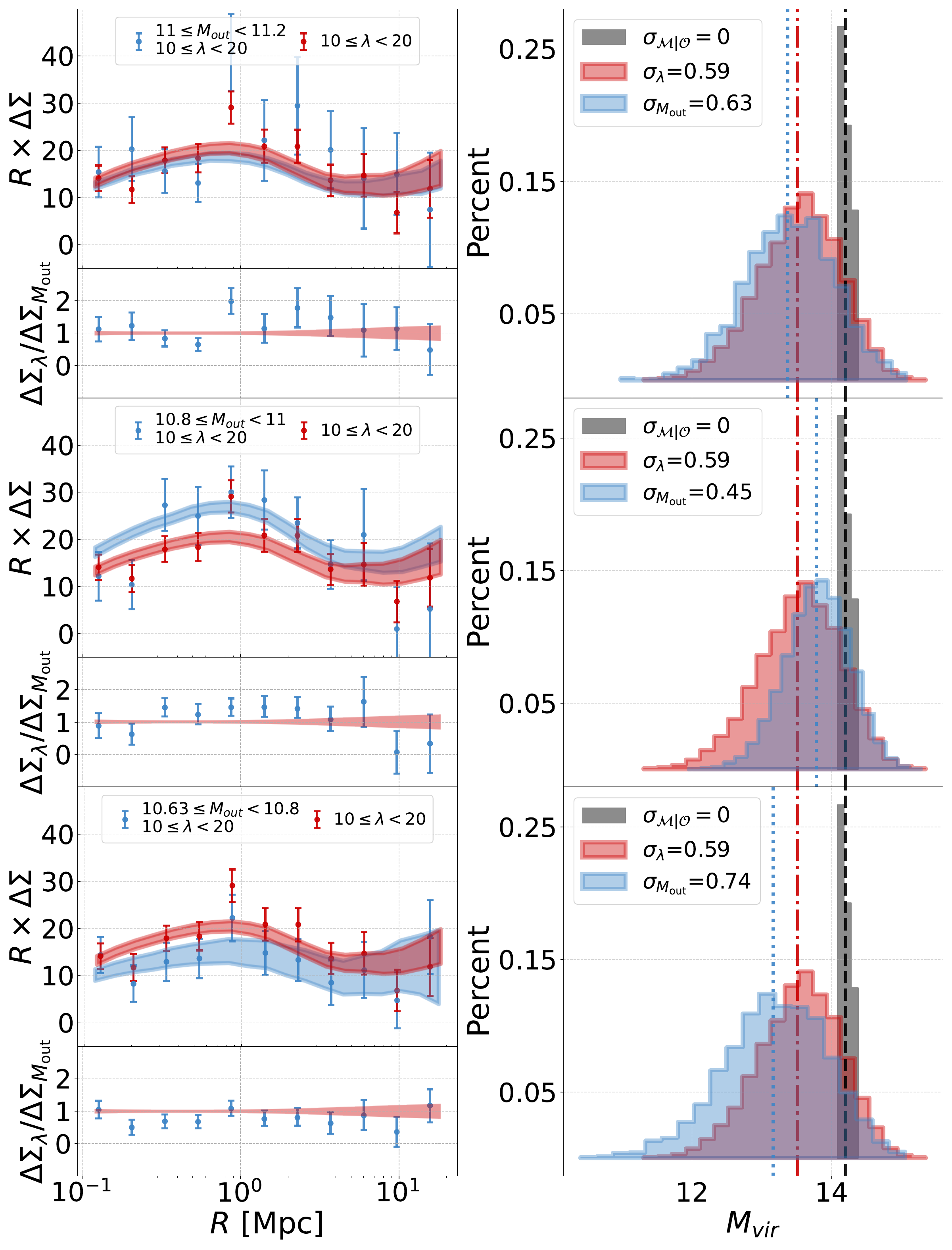}
    \includegraphics[width=.49\textwidth]{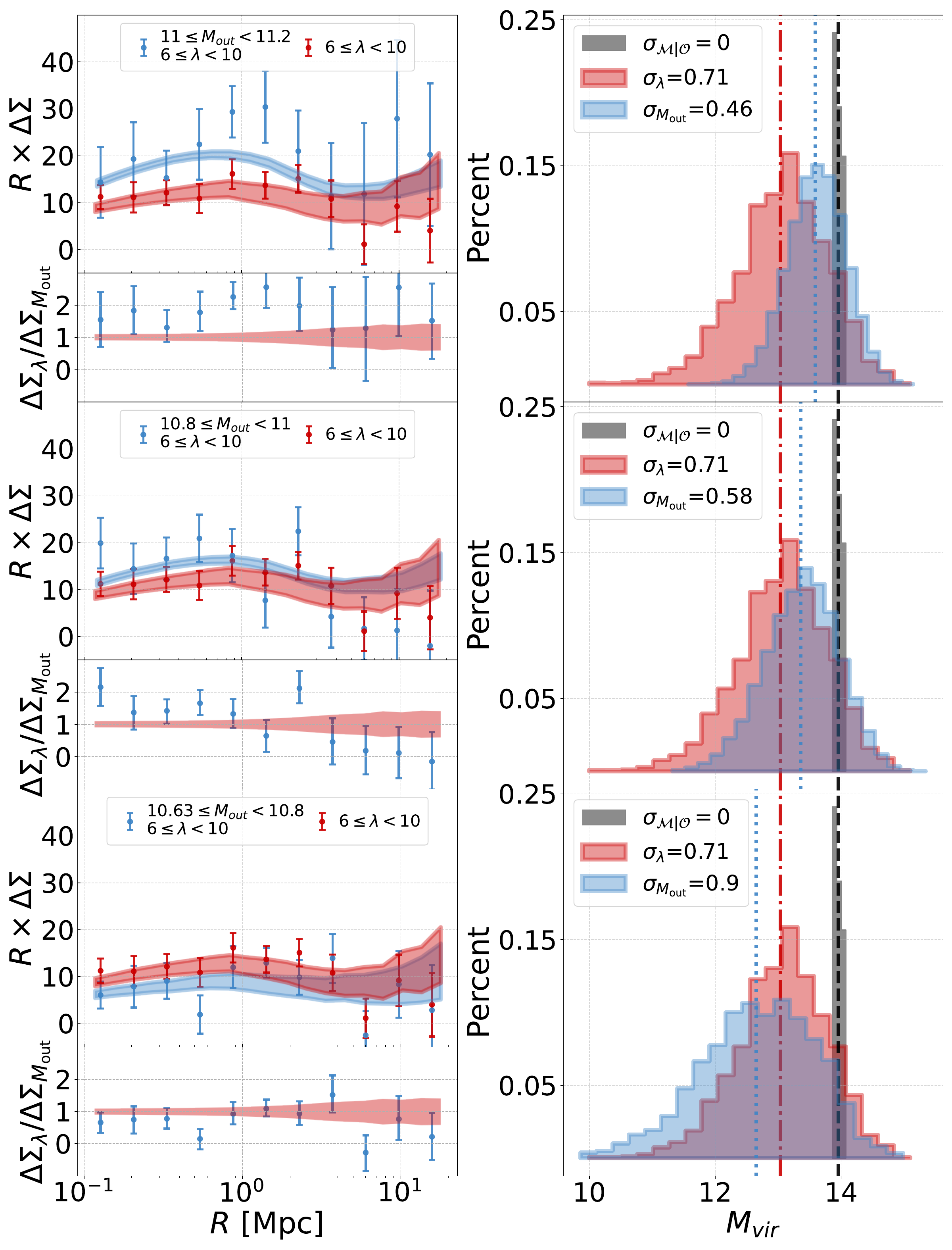}
    \caption{Measured $\Delta \Sigma$ signal and best-fit profiles for all clusters in $\lambda$ bin 3 (left column) and 4 (right column) and the measured $\Delta \Sigma$ signal and best-fit profiles when subdividing those $\lambda$ bins by $\mout$.  Each row corresponds to a different $\mout$ bin.  Clusters in each $\lambda$ bin are shown in red and clusters in each $\mout$ bin are shown in blue.  The lower left panel in each of the six plots in the figure show the ratio between the two $\Delta \Sigma$ profiles to highlight differences, and the histograms in the right panels show the $M_{\textrm{vir}}$ distribution within our simulations that generates the model, with grey being the distribution for a perfect tracer with $\sigma=0$.}
    \label{fig:subdiv-lambda}
    \end{center}
\end{figure*}

First, we present our results for how $\mout$ scatters within a $\lambda$ selection.  Our results for $10 \leq \lambda < 20$ (left column) and  $6 \leq \lambda < 10$ sample (right column) are shown in \autoref{fig:subdiv-lambda}.  Overall, there is consistency between best-fit $\Delta \Sigma$ amplitudes and $\sigma$.  In the data points for the left and right column, we find that the massive galaxies in $\smout$ that scatter to higher and lower $\mout$ have a higher and lower signal.  This is also reflected in the small difference in amplitude seen at radial scales below $10\Mpc$.  We see a clear difference in the top right figure containing clusters with $6 \leq \lambda < 10$ and $11 \leq \mout < 11.2$.   These clusters that scatter to a higher than expected $\mout$ have a signal consistent with a higher mass and lower scatter.  A similar trend is seen for clusters that scatter to a lower $\mout$ for the $10 \leq \lambda < 20$ sample.  This difference demonstrates two key points: that $\mout$ may contain independent information about the dark matter halo and that the $\mout$ measurements reported by $\smout$ correlate well with halo mass.

\begin{figure*}
    \begin{center}
    \includegraphics[width=.45\textwidth]{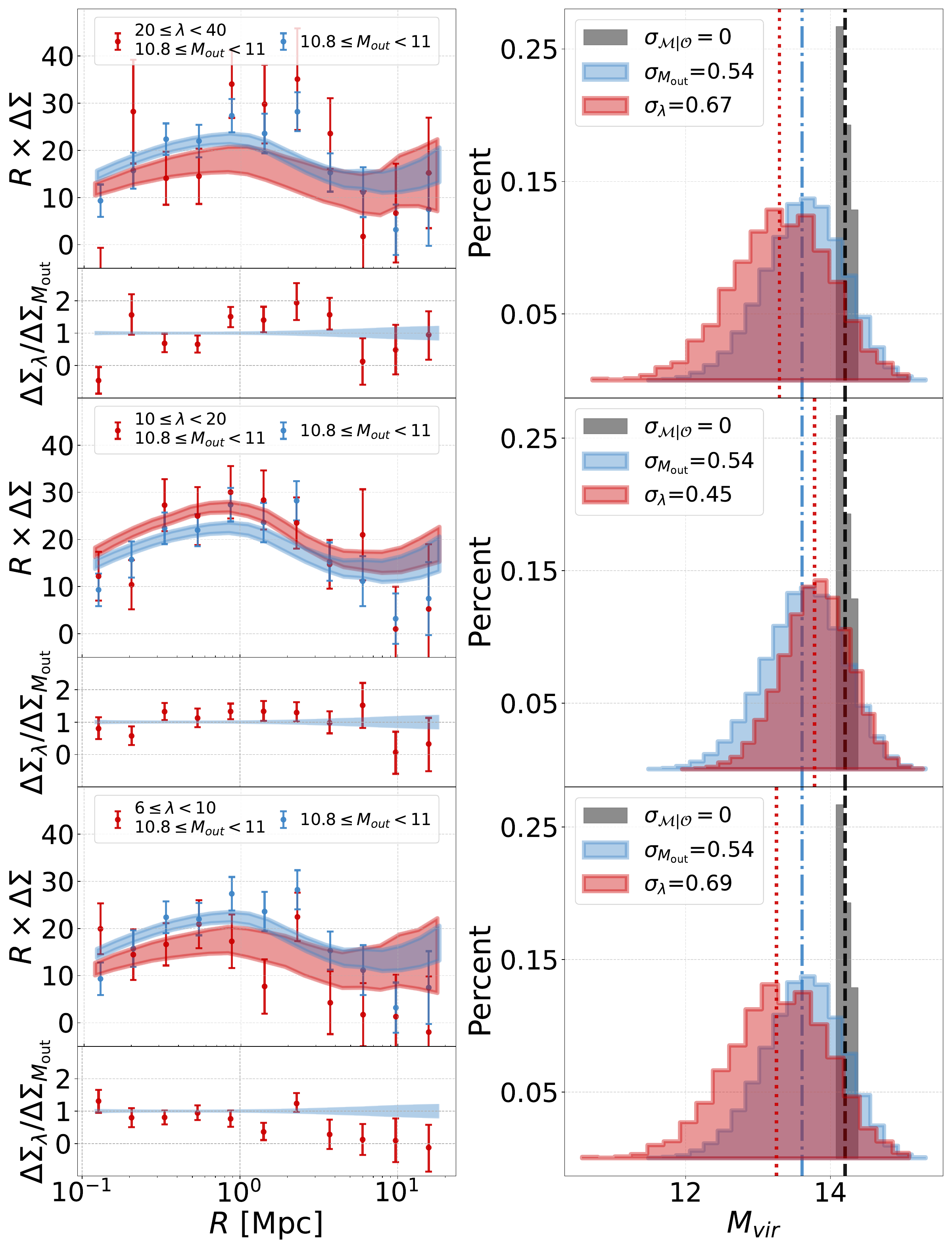}
    \includegraphics[width=.45\textwidth]{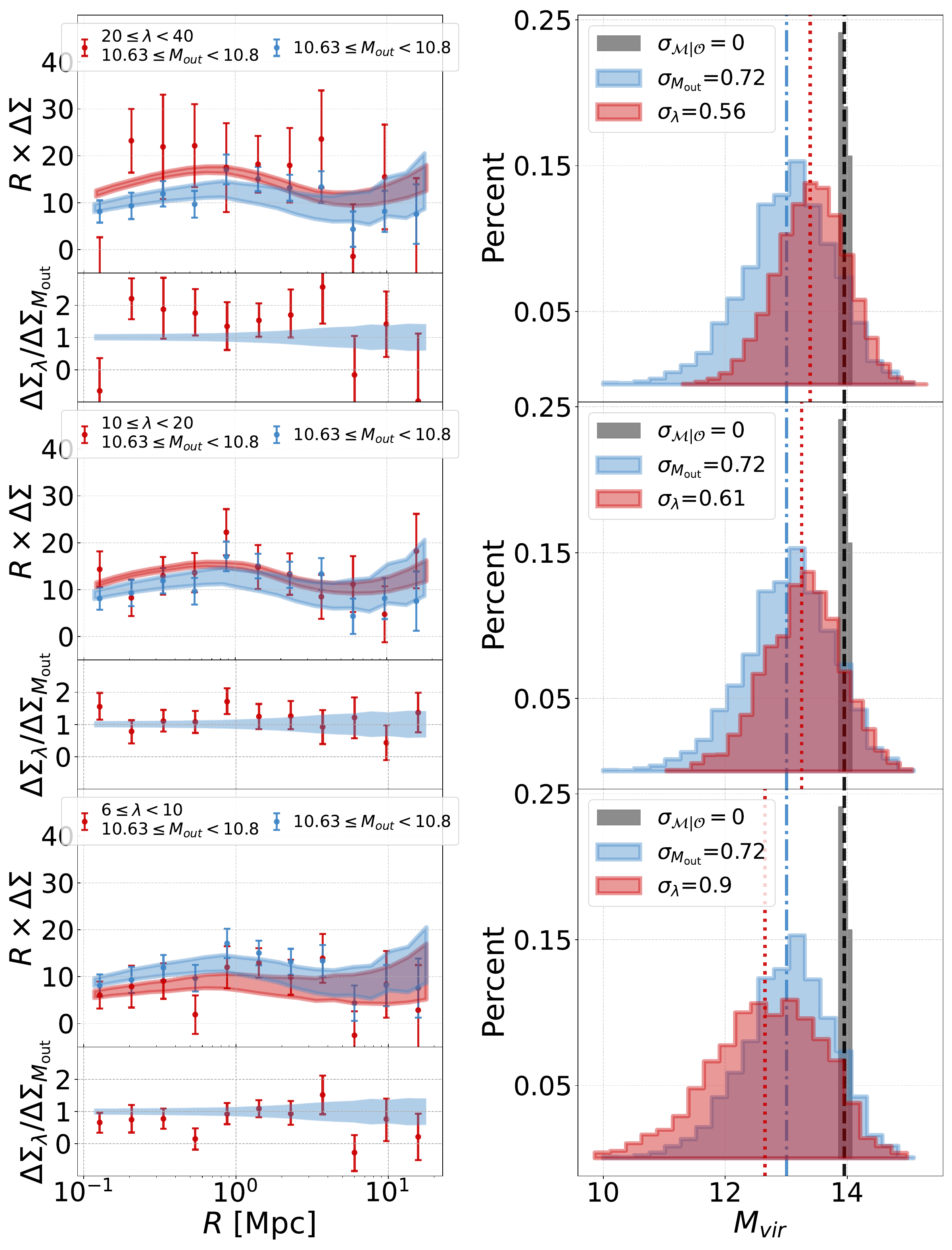}
    \caption{
    Measured $\Delta \Sigma$ signal and best-fit profiles for all clusters in $\mout$ bin 3 (left column) and 4 (right column) and the measured $\Delta \Sigma$ signal and best-fit profiles when subdividing those $\mout$ bins by $\lambda$.  Each row corresponds to a different $\lambda$ bin.  Clusters in each $\mout$ bin are shown in blue and clusters in each $\lambda$ bin are shown in red.  The lower left panel in each of the six plots in the figure show the ratio between the two $\Delta \Sigma$ profiles to highlight differences, and the histograms in the right panels show the $M_{\textrm{vir}}$ distribution within our simulations that generates the model, with grey being the distribution for a perfect tracer with $\sigma=0$.}
    \label{fig:subdiv-mass}
    \end{center}
\end{figure*}

Now, we turn to our results for how $\lambda$ scatters within a $\mout$ selection. We show the results for subdividing $\mout$ by $\lambda$ in \autoref{fig:subdiv-mass}.  Similarly to previous results, \redm clusters with $10.63 \leq \mout < 10.8$ that scatter to a higher $\lambda$ have a more peaked amplitude, and conversely, clusters which scatter to a lower $\lambda$ have a shallower amplitude.  The difference is less pronounced for $10.8 \leq \mout < 11$ and more similar to the results when subdividing $\lambda$ by $\mout$, where the best-fit profiles tend to remain consistent.  We draw the same conclusions: the $\lambda$ reported by \redm correlate well with cluster mass, and $\lambda$ carries independent information that $\mout$ selection alone does not capture.

These figures demonstrate that $\lambda$ and $\mout$ have consistent $\sigma$ with halo mass.  However, this consistency could be due to the relative size of our error bars and the shape of our best-fit models.  Looking specifically at the signal and not the best-fit model, there are discrepancies when either $\lambda$ or $\mout$ scatters higher or lower than the expected number density bin.  If this result holds with more data, then each tracer carries independent information about the underlying halo; a selection on $\lambda$ only, or $\mout$ only, would forgo available data about the amount of dark matter.  A combined cluster mass proxy of $\lambda$ and $\mout$ could have a lower scatter than each independently.  We comment on this more in the discussion.

\subsection{$\lambda$-$\mout$ scatter}
\label{sec:lm-scatter}

We find the log-normal scaling relation between $\mout$ and $\lambda$ using the fully Bayesian approach created by \citet{Kelly07} and implemented in \textsc{CluStR}\footnote{https://github.com/sweverett/CluStR} that allows for correlated measurement errors and intrinsic scatter in the regression.  We use errors reported by \redm for our uncertainties on $\lambda$ and errors in the \texttt{cmodel} stellar mass estimates for uncertainties on $\mout$. We find the following scaling relations and report our results in the form 
$$
\ln{y} = \alpha \ln \left(\frac{x}{x_{\textrm{piv}}}\right) + \beta, 
$$
where $\beta$ is the y-intercept, $\alpha$ is the slope of the scaling relation, and the pivot $x_{\textrm{piv}}$ is taken to be the median of our independent variable ($\mout$). 

\begin{figure*}
    \begin{center}
    \includegraphics[width=.49\textwidth]{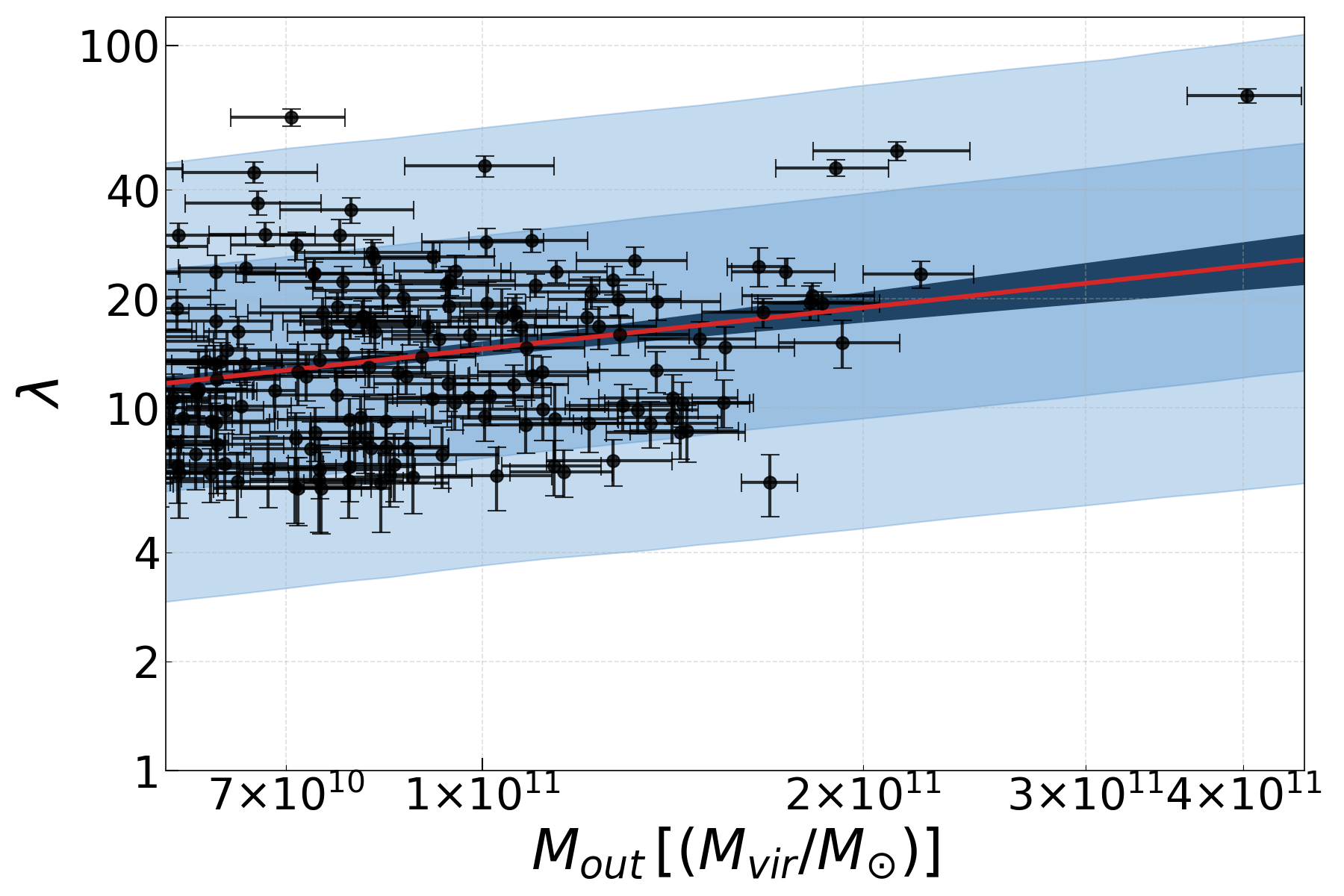}
    \includegraphics[width=.49\textwidth]{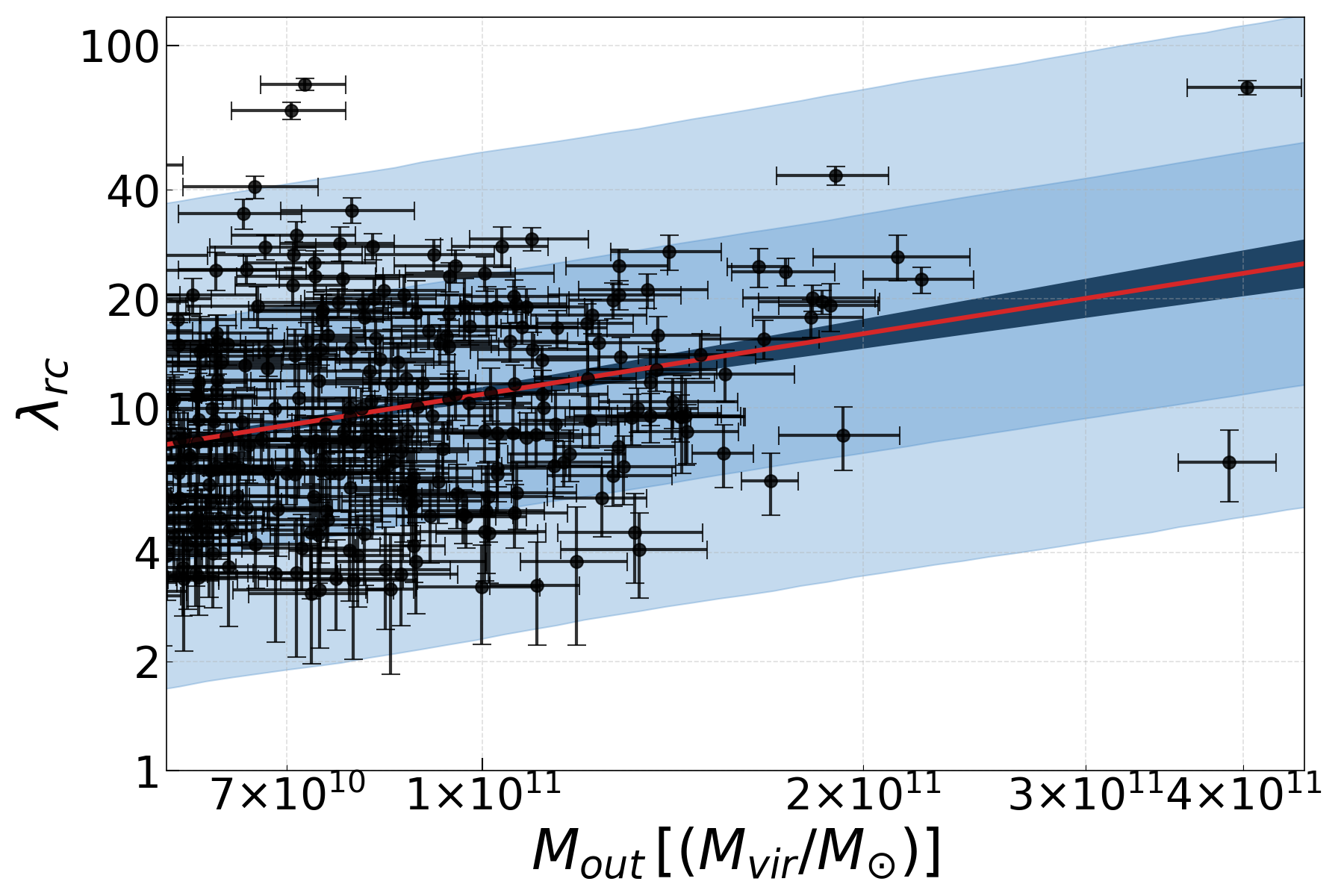}

    \caption{$\lambda$ - $\mout$ log-normal scaling relation from the matched \redm catalog (left) and \redm \texttt{runcat} catalog (right). The black dots are \redm members, and $\mout$ selected massive galaxies and their associated error bars.  The red line gives the best-fit scaling relation, and the black shaded region gives the $1\sigma$ uncertainty in the fit.  The blue-shaded region gives the $1\sigma$ intrinsic scatter between the two observables, and the light blue is the $2\sigma$ intrinsic scatter. For the left plot we find $\ln{\lambda} = (0.38 \pm 0.09) \ln \left(\frac{\mout}{M_{\textrm{pivot}}}\right) + (2.5 \pm 0.09)$ with $\sigma = 0.49 \pm 0.02$, where $M_{\textrm{pivot}} = 6.96 \times 10^{10}$. For the right plot we find $\ln{\lambda_{\textrm{rc}}} = (0.55 \pm 0.08) \ln \left(\frac{\mout}{M_{\textrm{pivot}}}\right) + (2.1 \pm 0.06)$ with $\sigma = 0.59 \pm 0.02$, where $M_{\textrm{pivot}} = 6.15 \times 10^{10}$.}
    \label{figure:lognormalscatter}
    \end{center}
\end{figure*}

The results of fitting all match categories with percolated $\mout$ galaxies removed is given by
\begin{equation}
\ln{\lambda} = (0.38 \pm 0.09) \ln \left(\frac{\mout}{M_{\textrm{pivot}}}\right) + (2.5 \pm 0.09)
\end{equation}
with  the scatter in $\lambda$ at fixed $\mout$ found to be $\sigma = 0.49 \pm 0.02$ and where $M_{\textrm{pivot}} = 6.96 \times 10^{10}$.  This fit is shown in the left panel of \autoref{figure:lognormalscatter}. We also fit the \texttt{runcat} catalog which contains measured $\lambda$ values at each $\mout$ galaxy, and find 
\begin{equation}
\ln{\lambda_{\textrm{rc}}} = (0.56 \pm 0.08) \ln \left(\frac{\mout}{M_{\textrm{pivot}}}\right) + (2.1 \pm 0.06)
\end{equation}
with the scatter in $\lambda_{\textrm{rc}}$ at fixed $\mout$ found to be $\sigma = 0.59 \pm 0.02$ and where $M_{\textrm{pivot}} = 6.15 \times 10^{10}$.  This fit is shown in the right panel of \autoref{figure:lognormalscatter}. Here the slope steepens somewhat and the scatter increases as is expected since we no longer have a strict $\lambda$ cut and are more fully sampling the scatter in richness with $\mout$.

We compare our derived scaling relation to \citet{Golden-Marx23}, where those authors quantify the stellar mass to halo mass scaling relation for different radial ranges to quantify the intra-cluster light (ICL) to halo mass scaling relation, using data from the DES and Atacama Cosmology Telescope Survey overlap \cite{Hilton21}.  This ICL is an alternative measurement of the \textit{ex-situ} stellar mass we are attempting to capture with our $50-100\kpc$ radial range. While those authors report the stellar mass to halo mass relation, we can qualitatively compare our findings, assuming that richness has additional scatter that reduces the scaling relation and increases the scatter in our findings. Those authors report the stellar mass to halo mass relation for total stellar mass within $50-100\kpc$ to be $\alpha = 0.297 \pm 0.136$ and $\sigma = 0.256 \pm 0.012$, where this is the intrinsic scatter in ICL stellar mass at fixed halo mass.  Additionally, \citet{Golden-Marx24} examines the same scaling relation in the $30-80\kpc$ range and finds $\alpha = 0.431 \pm 0.031$ and $\sigma = 0.211 \pm 0.004$.  Compared to our findings, we find a larger scatter of $\sigma = 0.49 \pm 0.02$.  This difference is as expected since the $\mout$ to $\lambda$ relation will increase scatter in our measured scaling relation. 

\section{Discussion}\label{sec:discussion}
We have used $\Delta \Sigma$ measurements on clusters selected by $\lambda$ and $\mout$ to better quantify the performance of $\mout$ as a halo mass proxy.  Our results demonstrate that $\mout$ and $\lambda$ have consistent scatter with $\mvir$ while having separate systematic effects and selection biases that influence purity at low halo mass.  We begin our discussion by examining these different systematics and speculate on potential calibration using spectroscopic data from DESI.  We also comment on the incidence of the CG-BLEND and SAT-BLEND failure modes.  Then, we assess the potential of $\mout$ in selecting clusters.  Throughout our analysis, we found features in the $\Delta \Sigma$ signal of $\lambda$ selected clusters that were not well fit by a model with only log-normal scatter, and we consider the possibility of avoiding this with a $\mout$ selection.  We conclude by discussing combining $\lambda$ and $\mout$ into a single tracer with lower scatter.

\subsection{The Impact of Satellites and Mis-centering}
\label{subsec:disc-satellites}
The N-MATCH-NC, N-MATCH, and MATCH-NC categories are all cases where \redm could have selected the wrong cluster CG, and the matched $\mout$ galaxy could be the true cluster center.  We combined these categories for this reason into the \textit{RM Satellite} column in \autoref{table:category_stats}.  In the worst-case scenario, \redm is always picking the wrong central galaxy.  This column is, therefore, some measure of an upper limit on the amount of mis-centering present in $\srm$. We found a matched RM Satellite for $26 \pm 3 \%$ for all $\srm$ clusters, and $53 \pm 7 \%$ if we consider only $\lambda > 20$ clusters.  However, this mostly comes from lower mass satellites, as evidenced by the low fraction of N-MATCH and N-MATCH-NC clusters above $\lambda > 20$ and our visual inspection of the N-MATCH category.  Mis-centering in \redm catalogs has been studied in the DES using X-ray emission data. In the DES Y1 \redm catalog, $81^{+11}_{-8}\%$ of clusters were found to be well centered \citep{Zhang19}.  In the DES Y3 \redm catalog, this rose to $87 \pm 4\%$  of clusters being well centered \citep{Kelly24}.  We note that the mis-centering studies were carried out on the \textsc{IDL} version, whereas we use the \textsc{Python} version. However, we do not expect a difference in centering performance between the two versions as both algorithms use the same centering model and run on the same underlying galaxy catalog.  Our upper bound is quite a bit larger, which is logically consistent as it is unlikely that \redm is always incorrect in the N-MATCH, N-MATCH-NC, and MATCH-NC cases.  

The occurrence of satellite galaxies within $\smout$ is related to the total number of multiple matches of $\mout$ galaxies with the same redMaPPer cluster, specifically (N-MATCH-NC)-1 + (N-MATCH)-1, and we reported this in the \textit{Duplicate CL Match} column of \autoref{table:category_stats}.  If the measurements of $\mout$ had no failures and the redMaPPer membership determinations were perfect, the number of duplicates would represent a lower limit on the satellite fraction in $\smout$. Of the 63 \redm clusters in N-MATCH, N-MATCH-NC, and MATCH-NC, we have 46 cases where both the cluster and the matched $\mout$ galaxies have a spec-z.  We find only one of these galaxies has a velocity offset $v_{\textrm{off}} > 3,000 \kms$ indicating that it is not a cluster member; this result implies the duplicate fraction does not have large contributions from interloping galaxies.  We took the clusters in these categories and measured the $\Delta \Sigma$ profile at the matched $\mout$ galaxy and at the \redm CG.  We find no difference in amplitude, and present these findings in \autoref{subsec:redm-centering}. These results indicate \redm centering and membership assignment within the sample is mostly accurate, modulo a few clusters.  However, failure modes in detecting massive galaxies in $\mout$ and $\lambda$ mean that, while indicative, we can not strictly interpret the duplicate fraction as a limit on the satellite fraction.  
We find $16 \pm 2\%$ of $\smout$ galaxies match to a previously matched \redm cluster, and for $\mout \geq 11$, this number is $3 \pm 2\%$.  Comparing to previous work by \citet{Huang2022}, those authors find a satellite fraction between $5-10\%$ for galaxies with $\log_{10}{M_{\star}/M_{\odot}} > 11.7$.  Our $\mout \geq 11$ cut shares the same number density as a $\log_{10}{M_\star/M_\odot} > 11.77$  cut, making our fraction roughly comparable.  We cannot state whether we are consistent because of the CG-BLEND and SAT-BLEND failure modes.  The next step would be to identify satellites in $\mout$ catalogs. However, recent work has also shown that cosmology from massive galaxies is robust to contamination from satellite galaxies \citep{Xhakaj23}.  

The imminent release of spectra from the DESI collaboration may provide a straightforward calibration for satellites within an $\mout$ selected cluster catalog.  Assuming this is true, this could significantly improve the purity of an $\mout$ catalog at low halo masses.  Because of the nature of the HMF, this would greatly increase the statistical constraining power of cluster cosmology.  However, the same cannot be said for a $\lambda$ selected cluster catalog.  There are systematic effects in addition to mis-centering, such as projection effects, which are possible to constrain using spectroscopic data in aggregate in some mass and redshift ranges \citep{Myles2021}, but not on an individual cluster level, as this would require spectra for all potential cluster members. In DESI, fiber collisions limit the spectroscopic redshifts that are obtainable in dense regions, \citep{Yang19} leading to incomplete coverage of cluster members even at relatively low redshifts. In addition, the magnitude of member galaxies contributing to the redMaPPer richness at higher redshifts make complete spectroscopic follow up of a large number of clusters prohibitively expensive. For example, for a LSST cluster cosmology analysis, selection effects need to be understood out to $z \sim 1$, and since projection effects are richness, redshift, and galaxy luminosity dependent, we would need to sample these in many bins and down to faint galaxy luminosities. So while it is possible in theory to obtain the needed spectra, it is not very practical, and only needing to understand the selection of the most massive galaxy is very appealing.

\subsection{Cluster Finding with $\mout$}

Our results highlight possible improvements to cluster finding using $\mout$ as a mass proxy.  In \autoref{subsec:mamiss}, we demonstrate that clusters that are only detected in one sample tend to be poor tracers of halo mass and that \redm tends to detect more of these clusters at low richness than $\mout$ at low mass, though we note that these conclusions are primarily limited to $\lambda<20$ where our sample sizes are sufficient to sample the lensing profiles.  Including $\lambda < 20$ clusters in cluster cosmology may not be straightforward. Recent work using spectroscopy and SZ data point to projection biases on observed richness and contamination that grow at lower richnesses \cite{Myles2021,Grandis21}, and simulations show that the selection biases due to line-of-sight structure which lead to biases in cluster weak lensing profiles also depend on richness \cite{Wu22}; however these effects remain poorly constrained due to the limited data for low richness clusters and limitation in the simulations. %Recent work by \citet{Wu22, Salcedo23} demonstrates that un-modelled selection biases and systematic effects exist specifically at low richness in \redm cluster catalogs. 
 However, with spectroscopic data from DESI, improving $\mout$ purity may be much more straightforward, exponentially increasing the number counts on the cluster cosmology data vector. This increase in halos consequently decreases statistical uncertainties on cosmological parameters \citep{Xhakaj23}.  

The findings in \autoref{subsec:mamiss} and \autoref{subsec:result-number-density} also show that $\lambda$ and $\mout$ have consistent scatter and similar trend with $\mvir$.  Consistent scatter and $\mvir$ dependence imply that $\mout$ has $\sigma$ sufficient to be an effective cluster finder since the limiting factor in cluster cosmology lies in the weak lensing mass calibration, and not in the scatter of $\lambda$ \citep{DESY1-clusters}.  %Furthermore, \citep{Huang2022} demonstrates in 6.3 that the presence of correlated structure along the line of sight manifests in the $\Delta \Sigma$ signal as features not well fit by a log-normal relation with constant Gaussian scatter.  These projection effects are notoriously difficult to calibrate \citep{Costanzi19}. The lower $\chi^2$ reported in \autoref{subsec:chi-sq} for the $\mout$ model suggests that these projection effects are not present in the $\mout$ data vector.

As previously mentioned a significant difficulty in using $\lambda$ as a proxy lies in the ability to correct for projection effects which bias the observed richness \cite{Myles2021, Costanzi19} and introduce selection biases which in turn bias the weak lensing signal compared to a purely mass-selected sample \cite{DESY1-clusters,Sunayama2020,Wu22}.  Outer stellar mass is expected to be free from projection effects due to line-of-sight structure as the selection is based on a single galaxy rather than a population of galaxies.  Similar to our findings in this paper, \citet{Huang2022} demonstrates that the $\Delta \Sigma$ signal for an $\mout$-selected sample is well fit by the expected profile for a simple mass-observable relation with log-normal scatter, while a richness-selected sample shows features in the $\Delta \Sigma$ profile not captured by this simple model.

Selection and projection effects for redMaPPer are typically calibrated for using simulations, but even state-of-the-art simulations fail to achieve the correct number density of galaxies within galaxy clusters. Therefore, conclusions drawn from these simulations are limited \citep{DeRose19}.  \redm additionally requires simulations with accurate galaxy colors.  By cluster finding with $\mout$, it may be possible to sidestep these issues as $\mout$ has different selection biases that may not be as difficult to model and calibrate.  However, we emphasize that there will be other and potentially new systematic effects, like the presence of satellites that require calibration in $\mout$ selection. We suspect these may be more tractable than the $\lambda$ selection.  At higher redshifts, beyond those currently used for optical cluster cosmology, we expect the ICL to be less prominent as it has had less time to build up, and this might effect the robustness of $\mout$ selection at high redshifts. 

Lastly, with the imminent arrival of data from stage IV experiments such as LSST and DESI, the quality and quantity of stellar mass measurements will improve drastically \citep{LSST19, DESI16}.  $\mout$ is well suited to capitalize on these new data sets.

\subsection{Easier mass modelling with $\mout$?}
When selecting samples by $\lambda$ and measuring $\Delta \Sigma$, we found features that could not be well fit by a log-linear relationship with constant scatter, evidenced by the value of the $\chi^2$ for the best-fit model.  Specifically, these features show up in the red model shown on the left-hand side of \autoref{fig:scp_bin_comp} and in the purple model shown in the top-right of \autoref{fig:mnm-bin}. We interpret these features as systematic biases introduced by the richness selection, as the log-normal scatter only model fails to capture them. Not only was this feature also identified by \citet{Huang2022}, but there have also been studies showing that systematics such as mis-centering \citep{McClintock19, Murata19} and projection effects \citep{Sunayama2020} can increase $\Delta \Sigma$ at this radial scale for red-sequence based cluster finders \citep{Wu22, DESY1-clusters}.  Disentangling this systematic and calibrating for it remains to be done.

The impact of difficult-to-fit features in a cluster $\Delta \Sigma$ model can have far-reaching effects, directly impacting the ability to calibrate cluster masses in cosmology analyses.  \citet{DESY1-clusters} suspects the discrepancy in their mass estimation resides precisely in the modeling of the weak lensing signal.  We do not see this bump in the $\Delta \Sigma$ signals from the $\mout$ selected samples, suggesting that they can be well fit by a log-normal relation with constant Gaussian scatter.  Given its less complex systematic effects, we argue that $\mout$ selection could improve cluster mass calibration.

The bump in the $\Delta \Sigma$ signals for $\lambda$ selected samples could also affect the $\chi^2$ for our best-fit models.  We generated these for our observables using a log normal model with constant scatter. Therefore, the fit for a $\lambda$ selected sample's signal will be poor.  This poor fit could increase the amplitude and, consequently, artificially decrease the inferred scatter of the observable. 

\subsection{A combined $\lambda-\mout$ tracer}

We comment on combining $\lambda$ and $\mout$ into a single halo mass proxy and cluster finder.  Previously, we speculated on the ability of $\mout$ to probe to lower halo mass in cluster cosmology.  However, combining the two proxies may be possible without incorporating more data sets.  As discussed, our results in \autoref{fig:mnm-bin} and \autoref{table:category_stats} demonstrate that purity is a concern for $\srm$ at low $\lambda$ and $\smout$ at low $\mout$.  However, the contamination that drives the loss in purity is from unrelated systematics in the selection involved in each mass proxy.  Because these two systematics are independent, we postulate that the overall purity of a combined $\lambda-\mout$ cluster sample may be higher at lower halo masses. 

Our results in \autoref{subsec:result-number-density} may also point to a lower $\sigma$ when combining the two proxies, though here sample size means that currently this conclusion is limited to $\lambda<20$.  Because the model fit may be poor primarily due to introduced systematic biases in the $\Delta \Sigma$ signal, it is worthwhile to examine and consider the difference in the individual data points rather than only the best-fit models, which can be seen in both \autoref{fig:subdiv-lambda} and \autoref{fig:subdiv-mass}.  Thus, combining the two could produce a tracer with lower $\sigma$ than either alone.  Similarly, \citet{Golden-Marx23, Zhang24, Golden-Marx24} and \citet{Sampaio21} find that diffuse intracluster light traces cluster mass and could be used in combination with other optical mass tracers to improve mass estimates.

Further work is needed to combine these tracers into a single mass proxy.  As we have shown, selecting on a combination of the two is possible.  We performed this by first running \redm in \texttt{runcat} (forced) mode on all $\mout$ galaxies without percolation, rank ordering by $\mout$, and then running \redm again with a percolation cut on $\mout$.  The next step would be to combine $\mout$ and $\lambda$ into a single cluster tracer. We do not include this in the scope of this work, but one could use this new proxy to derive new mass-observable relations and cosmology.

\section{Summary and Conclusions}\label{sec:conclusions}

In this paper, we compared the performance of the $\mout$ cluster proxy to \redm richness, $\lambda$.  We used the DES Y3 \redm cluster catalog and a catalog of massive galaxies selected by $\mout$ in the HSC Survey S16A data release.  We matched and categorized cluster detections in both samples to better understand the difference in selection between each tracer.  We used the measured $\Delta \Sigma$ signal and best-fit model to estimate the scatter in each observable.  We used this estimator to compare the magnitude of $\sigma$ for $\lambda$ and $\mout$. We performed this analysis on three samples to draw conclusions about the selection and scatter of each mass proxy. The results of our tests were as follows:

\begin{itemize}
    \item $\mout$ and $\lambda$ have consistent scatter with halo mass, and similar $M_{\textrm{halo}}$ dependence across the full $M_{\textrm{halo}}$ range probed here.
    \item The amount of satellites in $\smout$ decreases drastically above $\mout > 11$ (equivalent to $\lambda > 20$), but increases at lower $\mout$.
    \item Clusters detected by only one mass proxy tend to have a higher $\sigma$, and \redm tends to detect more single-detection clusters at low $\lambda$. 
    \item $\lambda$ selected samples appear to introduce features not well fit by a log-normal scatter only model into their $\Delta \Sigma$ measurements, manifested as a bump in the signal around $R=1\Mpc$, making it more difficult to model accurately.  $\mout$ selected samples do not introduce these systematics and are well-fit by a simple log-normal and constant Gaussian scatter model.
    \item Blending due to multiple cores is an important systematic in $\mout$ measurement that must be accounted for.
\end{itemize}

We discussed how introducing spectroscopic redshifts from DESI could calibrate the satellite fraction in $\mout$ selected cluster samples and enable including lower halo masses in cosmology.  We also found evidence that both $\lambda$ and $\mout$ contain independent information about the amount of dark matter. Combining the two tracers could lead to a single mass proxy with lower $\sigma$ and increased purity at lower mass. Lastly, we fit a log-normal scaling relation to the matched \redm and HSC clusters, and the \redm catalog is forced to run at each $\mout$ selected cluster with $\mout > 10.63$.  We found the scaling relations to be $\ln{\lambda} = (0.38 \pm 0.09) \ln \left(\frac{\mout}{M_{\textrm{pivot}}}\right) + (2.5 \pm 0.09)$ with $\sigma = 0.49 \pm 0.02$ and where $M_{\textrm{pivot}} = 6.96 \times 10^{10}$ and $\ln{\lambda_\textrm{rc}} = (0.55 \pm 0.08) \ln \left(\frac{\mout}{M_{\textrm{pivot}}}\right) + (2.1 \pm 0.06)$ with $\sigma = 0.59 \pm 0.02$ and where $M_{\textrm{pivot}} = 6.15 \times 10^{10}$.

One of the significant features of $\lambda$ that makes it a desirable proxy is its low scatter with halo mass.  Our results suggest that this is also a feature of $\mout$, with the added benefit of easier to model $\Delta \Sigma$ profiles.  Suppose one combines the two mass proxies into a single cluster finder. In that case, it opens up the possibility of (i) a lower scatter mass proxy overall, (ii) a cluster selection that may be easier to calibrate, and (iii) including lower mass clusters than what is currently possible in cosmological analyses.  Ultimately, systematics and selection biases in $\lambda$ selected cluster samples hinder the constraining power of cluster abundances on cosmological parameters such as $S_8$.  This hindrance further motivates us to investigate $\mout$ as an individual cluster finder and better understand the correlation between $\lambda$ and $\mout$ to combine the two into a single mass proxy. 

\section*{Contribution Statement}
MK performed the main analyses and drafted the manuscript.  TJ and AL supervised the project including project planning and direction, methodologies and analyses used, and contributions to the writing. SH provided data and modeling codes for the project and advised on the analysis methods. ER provided the original redMaPPer catalog and advice on running redMaPPer.  SH and JL provided significant feedback on the draft.  SE and PK provided code and help for the scaling relation fits. CZ provided feedback during the data analysis.  YZ and TS were internal reviewers.

\section*{Acknowledgements}
Matthew Kwiecien thanks Eileen Connor for the countless insightful conversations that contributed to many of the ideas in this work. 

This work was supported by the U.S. Department of Energy, Office of Science, Office of High Energy Physics, under Award Number DE-SC0010107. This material is also based upon work supported by the National Science Foundation under Grant No. 2206696. SH acknowledges support from the National Natural Science Foundation of China Grant No. 12273015 and No. 12433003 and the China Crewed Space Program through its Space Application System. Funding for the DES Projects has been provided by the U.S. Department of Energy, the U.S. National Science Foundation, the Ministry of Science and Education of Spain, 
the Science and Technology Facilities Council of the United Kingdom, the Higher Education Funding Council for England, the National Center for Supercomputing 
Applications at the University of Illinois at Urbana-Champaign, the Kavli Institute of Cosmological Physics at the University of Chicago, 
the Center for Cosmology and Astro-Particle Physics at the Ohio State University,
the Mitchell Institute for Fundamental Physics and Astronomy at Texas A\&M University, Financiadora de Estudos e Projetos, 
Funda{\c c}{\~a}o Carlos Chagas Filho de Amparo {\`a} Pesquisa do Estado do Rio de Janeiro, Conselho Nacional de Desenvolvimento Cient{\'i}fico e Tecnol{\'o}gico and 
the Minist{\'e}rio da Ci{\^e}ncia, Tecnologia e Inova{\c c}{\~a}o, the Deutsche Forschungsgemeinschaft and the Collaborating Institutions in the Dark Energy Survey. 

The Collaborating Institutions are Argonne National Laboratory, the University of California at Santa Cruz, the University of Cambridge, Centro de Investigaciones Energ{\'e}ticas, 
Medioambientales y Tecnol{\'o}gicas-Madrid, the University of Chicago, University College London, the DES-Brazil Consortium, the University of Edinburgh, 
the Eidgen{\"o}ssische Technische Hochschule (ETH) Z{\"u}rich, 
Fermi National Accelerator Laboratory, the University of Illinois at Urbana-Champaign, the Institut de Ci{\`e}ncies de l'Espai (IEEC/CSIC), 
the Institut de F{\'i}sica d'Altes Energies, Lawrence Berkeley National Laboratory, the Ludwig-Maximilians Universit{\"a}t M{\"u}nchen and the associated Excellence Cluster Universe, 
the University of Michigan, NSF NOIRLab, the University of Nottingham, The Ohio State University, the University of Pennsylvania, the University of Portsmouth, 
SLAC National Accelerator Laboratory, Stanford University, the University of Sussex, Texas A\&M University, and the OzDES Membership Consortium.

Based in part on observations at NSF Cerro Tololo Inter-American Observatory at NSF NOIRLab (NOIRLab Prop. ID 2012B-0001; PI: J. Frieman), which is managed by the Association of Universities for Research in Astronomy (AURA) under a cooperative agreement with the National Science Foundation.

The DES data management system is supported by the National Science Foundation under Grant Numbers AST-1138766 and AST-1536171.
The DES participants from Spanish institutions are partially supported by MICINN under grants PID2021-123012, PID2021-128989 PID2022-141079, SEV-2016-0588, CEX2020-001058-M and CEX2020-001007-S, some of which include ERDF funds from the European Union. IFAE is partially funded by the CERCA program of the Generalitat de Catalunya.

We  acknowledge support from the Brazilian Instituto Nacional de Ci\^encia
e Tecnologia (INCT) do e-Universo (CNPq grant 465376/2014-2).

This manuscript has been authored by Fermi Research Alliance, LLC under Contract No. DE-AC02-07CH11359 with the U.S. Department of Energy, Office of Science, Office of High Energy Physics.

\begin{appendices}
\renewcommand{\thefigure}{\thesection.\arabic{figure}}
\setcounter{figure}{0}  

\section{Appendix}

\begin{figure}
    \begin{center}
    \includegraphics[width=.49\textwidth]{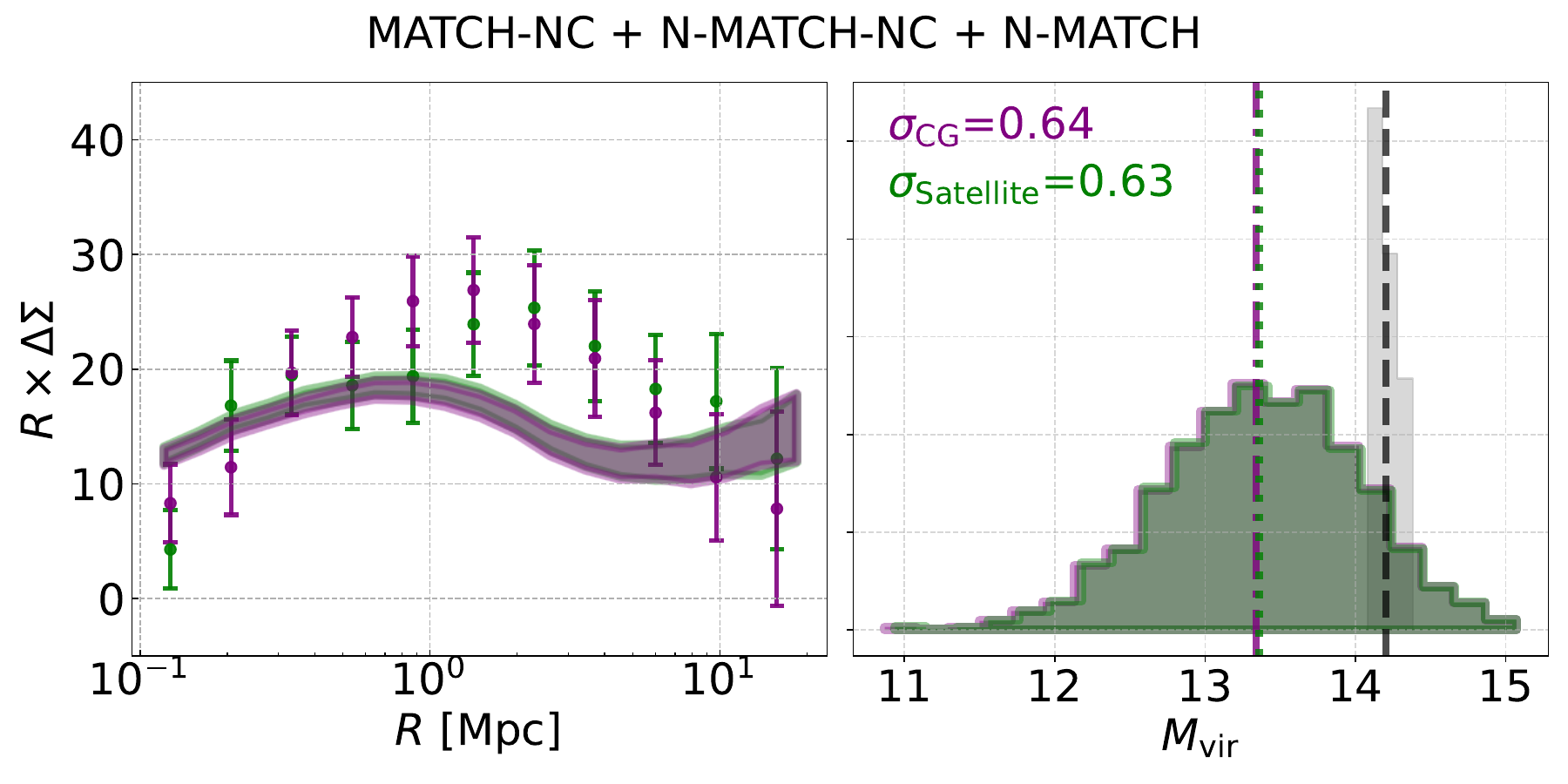}
    \includegraphics[width=.49\textwidth]{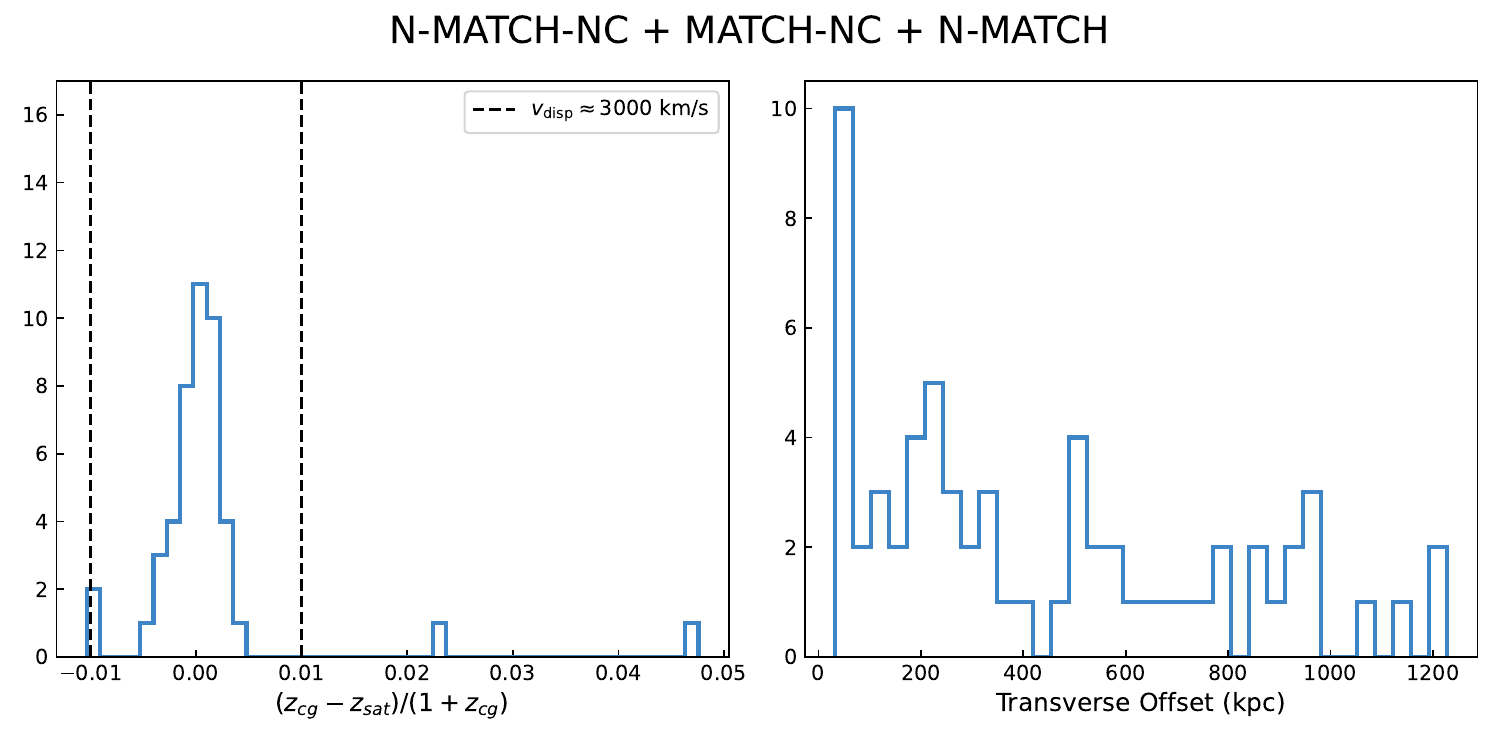}
    \caption{The results of our validation tests of the centering and membership probabilities in $\srm$ using clusters in the N-MATCH, N-MATCH-NC, and MATCH-NC categories. The top row shows the $\Delta \Sigma$ signal and associated $\sigma$ when treating the \redm CG (purple) as the cluster's center and then treating the matched \redm satellite (green) as the cluster's center. The bottom shows the line of sight and transverse physical offset between the \redm CG and the matched \redm satellite.  For the redshift offsets, we only include the cases where the CG and the \redm satellite have a spectroscopic redshift. We do not have the signal to see a difference between the \redm CG and satellite. However, we know that we have at most two \redm members that are not cluster members.}
    \label{app:match_nc_offsets}
    \end{center}
\end{figure}

\subsection{$\srm$ Centering and Membership Validation}\label{subsec:redm-centering}

We use the MATCH-NC, N-MATCH-NC, and N-MATCH categories to validate the centering and membership assignments in $\srm$.  We first validate the centering by comparing the $\Delta \Sigma$ profile at the \redm CG and then at the matched \redm satellites.  We show the results of this in the top row of \autoref{app:match_nc_offsets}.  A lower amplitude of the $\Delta \Sigma$ profile would suggest mis-centering \citep{McClintock19}.  Due to the size of our error bars, we cannot determine a difference between these $\Delta \Sigma$ models.  However, the lack of a significant difference would indicate that these are not all mis-centered.

Then, we test membership assignments by calculating the line of sight and transverse offsets between \redm CGs and the matched \redm Satellites in the cases where both have a spectroscopic redshift. We show our results in the bottom row of \autoref{app:match_nc_offsets}. Members with $v_{\textrm{disp}} > 3000 \kms$, or $|z_{\textrm{sat}} - z_{\textrm{cent}}| > 0.01$ can be assumed to be false \redm cluster members \citep{Rozo15, Wetzell22}. We find two clear spurious cluster members and two additional members close to this limit, which suggests that the membership assignments are accurate for the \redm clusters in these categories, except for a few galaxies.

\end{appendices}

\bibliographystyle{apsrev}
\bibliography{main}

\label{lastpage}
\end{document}